# Mechanical rotation via optical pumping of paramagnetic impurities

Pablo R. Zangara[1], Alexander Wood[3], Marcus W. Doherty[4], Carlos. A. Meriles[1,2,†]

[1]Dept. of Physics, CUNY-City College of New York, New York, NY 10031, USA. [2]CUNY-Graduate Center, New York, NY 10016, USA. [3]School of Physics, The University of Melbourne, Melbourne, Victoria 3010, Australia. [4]Laser Physics Centre, Research School of Physics and Engineering, Australian National University, Canberra, Australian Capital Territory 0200, Australia.

Hybrid quantum systems exhibiting coupled optical, spin, and mechanical degrees of freedom can serve as a platform for sensing, or as a bus to mediate interactions between qubits with disparate energy scales. These systems are also creating opportunities to test foundational ideas in quantum mechanics, including direct observations of the quantum regime in macroscopic objects. Here, we make use of angular momentum conservation to study the dynamics of a pair of paramagnetic centers featuring different spin numbers in the presence of a properly tuned external magnetic field. We examine the interplay between optical excitation, spin evolution, and mechanical motion, and theoretically show that in the presence of continuous optical illumination, inter-spin cross-relaxation must induce rigid rotation of the host crystal. The system dynamics is robust to scattering of spin-polarized phonons, a result we build on to show this form of angular momentum transfer should be observable using state-of-the-art torsional oscillators or trapped nanoparticles.

## I. INTRODUCTION

Growing applications in metrology and quantum information science are driving renewed interest in the interplay between spin and mechanical degrees of freedom, using one or the other as an interface to mediate quantized excitations[1,2]. Ingenious paths to controllably couple and manipulate spin and physical motion are also being explored as test beds for generating macroscopic quantum superposition and studying the boundaries between the quantum and classical worlds. For example, recent proposals on wave matter interferometry suggest the use of color centers in diamond as a handle to create translation[3] or rotation[4,5] superposition states in ~100 nm size particles. Conversion between spin and mechanical rotation also lies at the heart of the Einstein-de Haas and Barnett effects[6,7], long exploited as the preferred routes to determine the effective gyromagnetic ratio of charge carriers in ferromagnetic materials. Recent extensions have built on the higher sensitivity of torsional micro-cantilevers to investigate, for instance, engineered magnetic multi-layers[8], systems where the generation of a mechanical torque arises from domain wall displacements[9]. Other studies have examined torques generated by electron spin-flips in nanoscale systems[10], and magnetization tunneling in a single-molecule magnet coupled to a carbon nanotube resonator[11].

On a complementary front, active feedback and cavity-assisted schemes have been developed to gain control on the dynamics of optically trapped dielectric particles including both their center-of-mass motion and rotation[12,13]. Driving this effort is the race to attain high rotation speeds as a strategy to explore centrifugal forces and vacuum-friction effects[14]. Unlike translational degrees of freedom (evolving under harmonic oscillator forces and hence characterized by equidistant energy levels), rotational degrees of freedom have a non-linear energy spectrum and zero ground state energy, which can be exploited, e.g., to better study the superposition of rotational states in mesoscopic systems (the analog of persistent counter-propagating currents in a superconducting circuit), or for practical applications such as gyroscopy[15].

Thus far, all routes to driving particle rotation — both proposed and demonstrated — rely on the rotator's birefringence[16], or on the transfer of angular momentum from the light beam, assumed either circularly polarized or endowed of orbital angular momentum[17]. Here, we theoretically investigate an alternative form of opto-mechanics arising from a pair of interacting paramagnetic centers subject to light-induced spin pumping; an external magnetic field is adjusted so that the defects — assumed to have different spin numbers — can cross relax. For concreteness, we focus on the pair formed by a negatively-charged nitrogen-vacancy (NV) and a P1 center in diamond, though our ideas can be generalized to alternative pairs of defects, both in diamond and in other semiconductors such as SiC. We show that in the presence of continuous optical excitation, energy-conserving spin-flips between the NV and P1 lead to a net transfer of angular momentum from the spin pair to the lattice, both in the form of spin-polarized phonons and rigid rotation of the crystal as a whole, with the latter being dominant. We find that even in the absence of external friction the system gradually slows down to attain a pseudo-terminal velocity, which can be tuned by varying the applied magnetic field. With an eye on experiment, we discuss the more realistic case of particles hosting multiple paramagnetic defects, and show that rigid rotation can, in principle, generate entanglement between non-interacting pairs. This finding, however, should not be seen as a practical hurdle since the coupling rate — inversely proportional to the crystal moment of inertia — is exceedingly slow, meaning that the additive action of multiple defect pairs controlled via magnetic resonance techniques can serve as a handle to act on the particle rotational dynamics.

---

[†]E-mail: cmeriles@ccny.cuny.edu

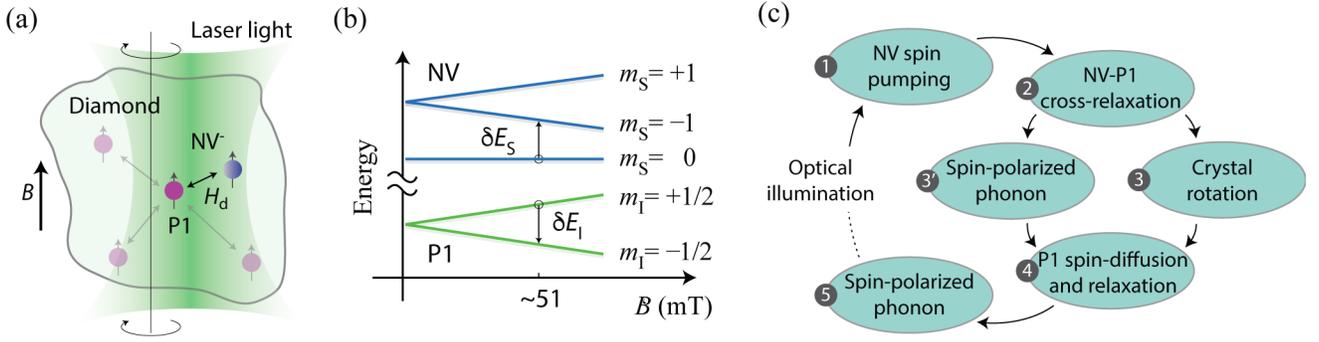

**Fig. 1: The interplay between optical spin pumping and the crystal's mechanical degrees of freedom.** (a) Schematics of a coupled NV-P1 pair in a diamond crystal. The P1 center interacts with other P1s farther removed from the NV. (b) Energy level diagram of the NV and P1 spins (top and bottom respectively. At ~51 mT the energies associated with each individual spin transition match, i.e., $\delta E_S \approx \delta E_I$. (c) Starting with NV spin optical initialization, the NV-P1 pair undergoes a cycle of cross relaxation and generation of spin-polarized phonons and rigid lattice rotation. The cycle completes with P1 spin diffusion and spin-lattice relaxation accompanied by the emission of spin-polarized phonons.

## II. ROTATIONALLY INVARIANT HAMILTONIAN

While the notion of spin-to-rotation conversion can, in principle, find various incarnations (see below), here we focus for concreteness on the spin pair formed by a P1 center (or neutral substitutional nitrogen impurity) and an NV center (in turn comprising a substitutional nitrogen adjacent to a vacancy). Figs. 1a and 1b respectively show a schematic and a simplified energy level diagram in the presence of an external magnetic field $B$ aligned with the symmetry axis of the NV. In its negatively charged state, the latter features a spin-1 ground state with a zero-field splitting of 2.87 GHz. Near 51 mT, the energy separation between the $m_S = 0$ and $m_S = -1$ ground state levels of the NV spin matches the Zeeman splitting between the $m_I = \pm 1/2$ levels of the P1. Continuous optical illumination (e.g., at 532 nm) preferentially pumps the NV spin into the $m_S = 0$ state, from where it subsequently transitions to the $m_S = -1$ state through dipolar-field-mediated cross-relaxation with the P1. In a typical type Ib diamond, the P1 concentration is comparatively higher, meaning that the polarization gained by the P1 proximal to the source NV can easily spin-diffuse to other, farther-removed defects. The end result is a one-directional spin-pumping process, from the NV to the ensemble of P1 centers, with the P1 steady-state polarization emerging from the interplay between the P1 concentration (defining the spin diffusion constant), and the defect's spin-lattice relaxation time. This process has already been investigated using optically-detected magnetic resonance (ODMR) both in the lab- and rotating-frames[18-21], and, more recently, has been exploited to induce high-levels of $^{13}$C spin polarization in diamond[22,23].

A closer inspection of the energy diagram in Fig. 1b indicates the above spin pumping process is deceivingly simple: While it is apparent that energy conservation can be ensured with the proper selection of the magnetic field, cross-relaxation of the NV-P1 pair entails a simultaneous flip to spin states with lower quantum projection numbers, thus leading to a net reduction of the total spin angular momentum by $2\hbar$. In other words, cross-polarization of the P1 spin requires the transfer of angular momentum to 'lattice' degrees of freedom, either through the generation of spin-polarized phonons[24,25], or the rigid rotation of the crystal (Fig. 1c).

To gain a more formal understanding, we first consider the case of a perfectly rigid solid (as we show later, a reasonable approximation for diamond). In this limit, the P1-NV virtual-atom pair can be thought of as forming a rigid diatomic molecule, featuring the crystal's moment of inertia $\mathcal{J}$. Correspondingly, we write the system Hamiltonian as

$$H = \Delta S_z^2 + \omega_0 S_z + \omega_0 I_z + H_d + \frac{L_z^2}{2\mathcal{J}}, \qquad (1)$$

where $\mathbf{S}$ ($\mathbf{I}$) is the NV (P1) vector spin operator, $\Delta$ is the NV zero field splitting, $\omega_0 \equiv |\gamma_e|B$ is the electron Zeeman frequency in the magnetic field $\mathbf{B}$ (assumed along the $z$-axis and parallel to the NV), $\gamma_e$ denotes the electron gyromagnetic ratio, $\mathbf{L}$ is the vector operator representing the crystal angular momentum, and we assume for simplicity the system can only rotate about the $z$-axis. In Eq. (1), $H_d$ expresses the NV–P1 dipolar interaction, here viewed as the coupling Hamiltonian $H_{s-r}$ between the spin pair and the (rigid) crystal rotation. To expose interconversion between spin and crystal rotation, we write

$$H_{s-r} \approx H_d = d_0(\mathbf{r})\delta_0 + d_1(\mathbf{r})\lambda_+\delta_{1-}$$
$$+ d_1^*(\mathbf{r})\lambda_-\delta_{1+} + d_2(\mathbf{r})\lambda_+^2 \delta_{2-} + d_2^*(\mathbf{r})\lambda_-^2 \delta_{2+}, \quad (2)$$

where $\delta_0 = S_z I_z - \frac{1}{4}(S_- I_+ + S_+ I_-)$, $\delta_{1\pm} = S_z I_\pm + S_\pm I_z$, $\delta_{2\pm} = S_\pm I_\pm$ are the two-spin operators in spherical tensor form. By the same token, we denote $d_0 = \frac{\alpha}{r^3}(1 - 3\cos^2\theta)$, $d_1 = -\frac{3\alpha}{2r^3}\sin\theta\cos\theta\, e^{i\varphi}$, $d_2 = -\frac{3\alpha}{4r^3}\sin^2\theta\, e^{2i\varphi}$, and $\lambda_\pm = e^{\pm i\phi}$. In the above expressions $\mathbf{r} = \mathbf{r}_{NV} - \mathbf{r}_{P1}$ is the inter-spin vector with polar and azimuthal angles $\theta$ and $\phi + \varphi$, respectively; the latter is expressed as the sum of the angle $\phi$ formed by the crystal relative to the laboratory frame and the (fixed) crystal frame azimuthal coordinate $\varphi$. Finally, $\alpha = \mu_0 \gamma_e^2/4\pi$, where $\mu_0$ denotes the vacuum permeability.

Assuming for now the regime where the crystal's rotational energy is smaller than the dipolar energy, we choose the external magnetic field so that $\omega_0 = \Delta/2$, the 'energy matching' condition required for NV–P1 cross-relaxation. Limiting our description to the spin subspace spanned by $|m_S, m_I\rangle = \{|0, +1/2\rangle, |-1, -1/2\rangle\}$, only the last two 'double-flip' terms in $H_d$ are (nearly) energy conserving, meaning that all first three contributions can be effectively truncated. In this limit, we rewrite the Hamiltonian as

$$H = d_2(\mathbf{r})\lambda_+^2 \delta_{2-} + d_2^*(\mathbf{r})\lambda_-^2 \delta_{2+} + \frac{L_z^2}{2\mathcal{J}}. \quad (3)$$

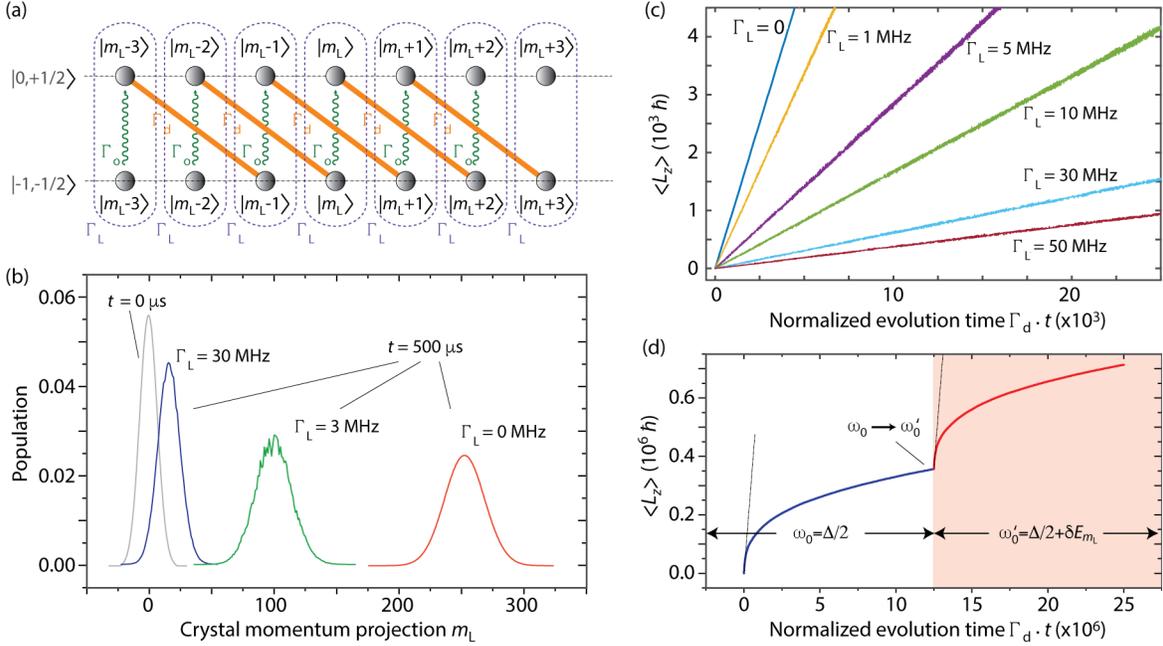

**Fig. 2: Modeling spin-crystal momentum conversion. (a)** 'Tight-binding' representation of the rotor pumping process: Upper and lower chains correspond to spin states $|0,+1/2\rangle$ and $|-1,-1/2\rangle$, respectively, while chain sites indicate rotational states $|m_L\rangle$. Starting from a state $|m_S, m_I, m_L\rangle$, the system evolution is governed by the dipolar rate $\Gamma_d$, the optical pumping rate $\Gamma_o$, and the rotor decoherence rate $\Gamma_L$. **(b)** Occupational probability of rotational states after evolution in the presence of continuous optical excitation for a fixed time interval $t = 500$ μs and various rotor dephasing rates $\Gamma_L$; the faint black trace indicates the population distribution assumed for $t = 0$. The dynamics is evaluated using the Trotter-Suzuki method (see Methods); each curve shows the result after 50 averages. **(c)** Mean angular momentum $\langle L_z \rangle$ as a function of the normalized evolution time $\Gamma_d t$ for some rotor decoherence rates $\Gamma_L$; the initial rotor state is that of (b). **(d)** Long-term evolution for the case $\Gamma_L = 0$. At sufficiently large rotational energies, the spin-crystal momentum transfer is inefficient and the system evolves towards a pseudo-terminal angular speed, whose value can be adjusted by shifting the magnetic field (shaded region of the plot). In (b), (c) and (d) we use $\Gamma_d = 0.5$ MHz, $\Gamma_o = 1$ MHz, and $\hbar/2\mathcal{J} = 10$ Hz.

Note that since $[L_z, \lambda_\pm] = \pm\hbar\lambda_\pm$, the $\lambda_\pm$ operators can be thought of as ladder operators to $L_z$. Therefore, Eq. (3) explicitly shows how angular momentum is conserved, namely, a net spin angular momentum loss from a double quantum flip is accompanied by a corresponding crystal angular momentum gain (and vice versa). The above dynamics is in strong contrast with the spin-conserving zero-quantum 'flip-flops' usually governing spin diffusion processes through terms of the form $d_0 I_\pm S_\mp$ in $H_d$. The key difference stems from the asymmetry created by the crystal field, acting on the NV but not the P1, and hence rendering flip-flop contributions to the Hamiltonian non-secular.

## III. SPIN-CRYSTAL ANGULAR MOMENTUM INTER-CONVERSION

To intuitively grasp the system dynamics in the presence of optical excitation, it is instructive to first consider the simplified case where the crystal — here seen as a free rotor — initially occupies a state $|m_L\rangle$ of mechanical angular momentum $m_L\hbar$, and a light pulse instantaneously projects the spin system into $|0,+1/2\rangle$ (we ignore for now the different initialization mechanisms in the NV and P1). Driven by the dipolar coupling, the NV–P1 pair evolves into $|-1,-1/2\rangle$ and, in so doing, changes the orbital part of the wavefunction into $|m_L + 2\rangle$. Reinitializing the spin system into $|0,+1/2\rangle$ repeats the process, but this time the rotor state evolves from $|m_L + 2\rangle$ into $|m_L + 4\rangle$, corresponding to buildup of the crystal angular momentum and hence to macroscopic physical rotation.

Under continuous optical excitation, this spin-induced rotational pumping can be best computed via the tight-binding representation of Fig. 2a where each linear chain corresponds to one of the two possible spin states and the site energies take values $E_{m_L} = m_L^2 \hbar^2/(2\mathcal{J})$. In the regime where the rotational energy is negligible (i.e., $E_{m_L} \ll \hbar|d_2|$, see below), the time evolution can be cast in terms of a series of inter-chain hops governed by unit time probabilities $\Gamma_d = \hbar|d_2|/(2\pi)$ and $\Gamma_o$, respectively representing the NV–P1 dipolar coupling and optical pumping rates. To realistically compute the system evolution, we must also take into account the rotor decoherence, which we model by imposing a dephasing rate $\Gamma_L$ on rotation states $|m_L\rangle$. Assuming a crystal with no net initial angular momentum (i.e., $\langle L_z \rangle(t = 0) = 0$) at some non-zero initial temperature $T_i$ (i.e., $\langle L_z^2 \rangle(t = 0) \propto T_i$), Fig. 2b compares the probability density of rotational states before and after a time interval $t$ of continuous optical excitation. Consistent with a crystal momentum gain, the initial distribution (faint black trace) invariably evolves to yield a net $\langle L_z \rangle$ — as reflected by the non-zero $\langle m_L \rangle$ — with significant momentum buildup even when $\Gamma_L \geq \Gamma_d$.

Fig. 2c shows the average angular momentum $\langle L_z \rangle$ as a function of time — expressed in units of the spin–crystal momentum transfer time $\Gamma_d^{-1}$ — for various rates of dissipation $\Gamma_L$. In all cases, we observe a linear growth, indicative of a constant torque on the crystal with value approximately proportional to $(\Gamma_d^2 + \Gamma_L^2)^{-1/2}$. This process, however, cannot be sustained indefinitely (even if $\Gamma_L = 0$) since, as $\langle L_z \rangle$ grows, so does the crystal's rotational energy (last term in Eq. (3)), whose increasingly larger energy steps $\delta E_{m_L} \equiv E_{m_L+2} - E_{m_L} = 2(m_L + 1)\hbar^2/\mathcal{I}$ become gradually comparable to the NV–P1 spin energy (i.e., $\delta E_{m_L} \sim 2\pi\hbar\Gamma_d$), thus slowing down the spin-crystal momentum conversion (Fig. 2d). Considering the extreme angular velocities demonstrated recently for optically driven nanoparticles[12,13] ($\gtrsim 2\pi \times 1$ GHz), this 'pseudo-terminal' regime — first reached for angular frequencies of order $\sim \Gamma_d$ — should be readily observable. Efficient rotational pumping, however, can be regained by changing the magnetic field so as to recover the 'energy matching' condition, i.e., $\Delta - 2\omega_0' + \delta E_{m_L} = 0$, where we use the prime to highlight the shift relative to the value $\omega_0 = \Delta/2$ at early stages (shaded half of Fig. 2d).

For future reference, it is possible to use Fermi's 'golden rule' to analytically calculate the rate of interconversion between NV spin polarization and crystal rotation. In the limit where $\Gamma_o \lesssim \Gamma_d$, we find (Supplemental Material, Section I)

$$\Gamma_{s-r} \approx \frac{4\pi^2 \eta \alpha^2}{5\hbar^2 r_{min}^3} \rho_{ss}(\Gamma_d, \Gamma_L) \approx \frac{\Gamma_d^2}{(\Gamma_d^2 + \Gamma_L^2)^{1/2}}, \quad (4)$$

where $\rho_{ss}(\Gamma_d, \Gamma_L)$ is a lineshape factor, and we assume a random distribution of NV–P1 pairs with number concentration $\eta$ and minimum separation $r_{min} = 1$ nm.

## IV. INTERACTION WITH SPIN-POLARIZED PHONONS

An alternative channel of momentum conservation is via phonons, recently shown[24,25] to carry an intrinsic "phonon-spin" angular momentum $\mathbf{L}' = \int_V d^3r' \rho\, \mathbf{u}(\mathbf{r}') \times \dot{\mathbf{u}}(\mathbf{r}')$, where $\mathbf{u}(\mathbf{r}')$ denotes the local lattice displacement vector, $\rho$ is the crystal density, and the integral extends over the crystal volume $V$. To explicitly describe the spin-phonon interaction we express the displacement vector as $\mathbf{u}(\mathbf{r}') = \sqrt{\hbar/(2\rho V)} \sum_{\mathbf{k},j} \mathbf{e}_{\mathbf{k},j} \exp(i\mathbf{k} \cdot \mathbf{r}') a_{\mathbf{k},j}/\sqrt{\omega_{\mathbf{k},j}} + $ H.c, where H.c denotes Hermitian conjugate, the sum extends over all wave-vectors $\mathbf{k}$ and (Cartesian) polarization branches $j$, $\mathbf{e}_{\mathbf{k},j}$ denotes the phonon polarization vector, $\omega_{\mathbf{k},j}$ is the phonon frequency, and we use the standard notation for the phonon creation and annihilation operators, respectively $a_{\mathbf{k},j}^\dagger$ and $a_{\mathbf{k},j}$. Replacing in the expression for $\mathbf{L}'$, one finds[26,27]

$$\mathbf{L}' = \hbar \sum_{\mathbf{k}} \frac{\mathbf{k}}{k}\left(a_{\mathbf{k}+}^\dagger a_{\mathbf{k}+} - a_{\mathbf{k}-}^\dagger a_{\mathbf{k}-}\right), \quad (5)$$

where $a_{\mathbf{k}\pm}^\dagger \equiv \mp\left(a_{\mathbf{k}1}^\dagger \pm i a_{\mathbf{k}2}^\dagger\right)/\sqrt{2}$, and the positive (negative) signs indicate left (right) circular polarization. Eq. (5) expresses the lattice spin angular momentum as the difference between populations of phonons, each carrying a unit $\hbar$ of angular momentum parallel or anti-parallel to the direction of propagation (i.e., positive or negative quantum of angular momentum, respectively).

To include the effect of spin-polarized phonons into the model, we expand $H_d$ to first order in the lattice displacements via the correspondence $\mathbf{r} \rightarrow \mathbf{r} + \delta\mathbf{r}$, where $\delta\mathbf{r} = \mathbf{u}(\mathbf{r}_{NV}) - \mathbf{u}(\mathbf{r}_{P1})$. After some algebra, we find (see Supplemental Material, Section II)

$$H_d \approx H_{s-r} + H_{s-p}, \quad (6)$$

where $H_{s-r}$ is the spin-rotation interaction derived above for the rigid rotator model (Eq. (2)), and

$$H_{s-p} \approx \sum_{\mathbf{k}} i\, \mathbf{k} \cdot \mathbf{r}\left(b_0 \pi_{\mathbf{k},0} \delta_0 + b_1 \pi_{\mathbf{k},1+}\delta_{1-} + b_1 \pi_{\mathbf{k},1-}\delta_{1+}\right). \quad (7)$$

Above we use the notation $\pi_{\mathbf{k},0} = \left[\frac{\hbar}{2\rho V \omega_\mathbf{k}}\right]\left(a_{\mathbf{k},z} - a_{\mathbf{k},z}^\dagger\right)$, $\pi_{\mathbf{k},1\pm} = \pm\left[\frac{\hbar}{2\rho V \omega_\mathbf{k}}\right]\left(a_{\mathbf{k},\pm}^\dagger + a_{\mathbf{k},\mp}\right)$, $b_0 = -\frac{3\alpha}{2r^4}\cos\theta\,(1 - 5\cos 2\theta)$, and $b_1 = -\frac{3\alpha}{16r^4}(3\cos\theta + 4\cos 3\theta)$. We also assume that $\mathbf{r}$ is small compared to the relevant phonon wavelengths, i.e., $\mathbf{k} \cdot \mathbf{r} \ll 1$ (see Supplemental Material, Section II). As in Eq. (3), the Hamiltonian of Eq. (7) makes rotational invariance explicit, this time through the interconversion of spin and phonon angular momentum. Unlike $H_{s-r}$, however, $H_{s-p}$ connects states differing, at most, by a *single* quantum of angular momentum. Since individual spin flips take place at a rate $\Gamma_{s-p}^{(1)}$ not greater than the inverse of the spin-lattice relaxation time $T_1^{(NV)} \sim T_1^{(P1)}$ (induced via $H_{s-p}$ or other spin-lattice relaxation processes[28], typically $\sim$1 ms at room temperature), we conclude $\Gamma_{s-p}^{(1)} \ll \Gamma_{s-r}$.

A possibility that must be considered separately, however, is one where double spin flips are allowed via second-order processes involving simultaneous absorption and emission of phonons. In this case, the combined spin-phonon system transitions from an initial state $|i_\mathbf{k}\rangle = |\ldots n_{\mathbf{k},-}, n_{\mathbf{k},+}, \ldots, 0, +1/2\rangle$ to a final state $|f_\mathbf{k}\rangle = |\ldots n_{\mathbf{k},-} - 1, n_{\mathbf{k},+} + 1, \ldots, -1, -1/2\rangle$. Here, the net spin of phonons with wave vector $\mathbf{k}$ — represented through spin-polarized phonon populations $n_{\mathbf{k},-}$ and $n_{\mathbf{k},+}$ — grows by two units of $\hbar$, hence compensating for the angular momentum change from NV–P1 spin cross-relaxation (last two quantum numbers in the kets). Note that other final states — involving, e.g., phonons with wave-vector different from the initial one — are forbidden, because spin cross-relaxation must conserve the total linear momentum and energy, i.e., $\mathbf{k}$ must remain unchanged.

To calculate the rate of spin-phonon momentum transfer via these second order pathways, we consider two types of mechanisms (Supplemental Material, Sections III and IV). In the first category, we group all off-resonance processes (i.e., $|\mathbf{k}| > |\mathbf{k}_0| \equiv \omega_0/c$, with $c$ denoting the speed of sound in diamond) where the transition from state $|i_\mathbf{k}\rangle$ to $|f_\mathbf{k}\rangle$ takes place via virtual states $|g_{\mathbf{k},j}\rangle$, $j = 1, 2$ involving an NV spin flip $|0\rangle \rightarrow |-1\rangle$ and the creation (annihilation) of a phonon with positive (negative) spin (Fig. 3a). The second group corresponds to resonant processes (i.e., $|\mathbf{k}| = |\mathbf{k}_0|$) involving an intermediate state $|g_{\mathbf{k}_0}\rangle$ with the same energy as $|i_{\mathbf{k}_0}\rangle$ or $|f_{\mathbf{k}_0}\rangle$ (Fig. 3b). Despite the massive majority of non-resonant relaxation channels, this second group of processes is more efficient in inducing spin-phonon conversion of angular momentum (at least at room temperature and below), mainly because phonon states with greater wave vectors quickly depopulate due to the stiffness of diamond (Supplemental Material, Section IV). After a lengthy calculation, we find the characteristic spin-phonon conversion

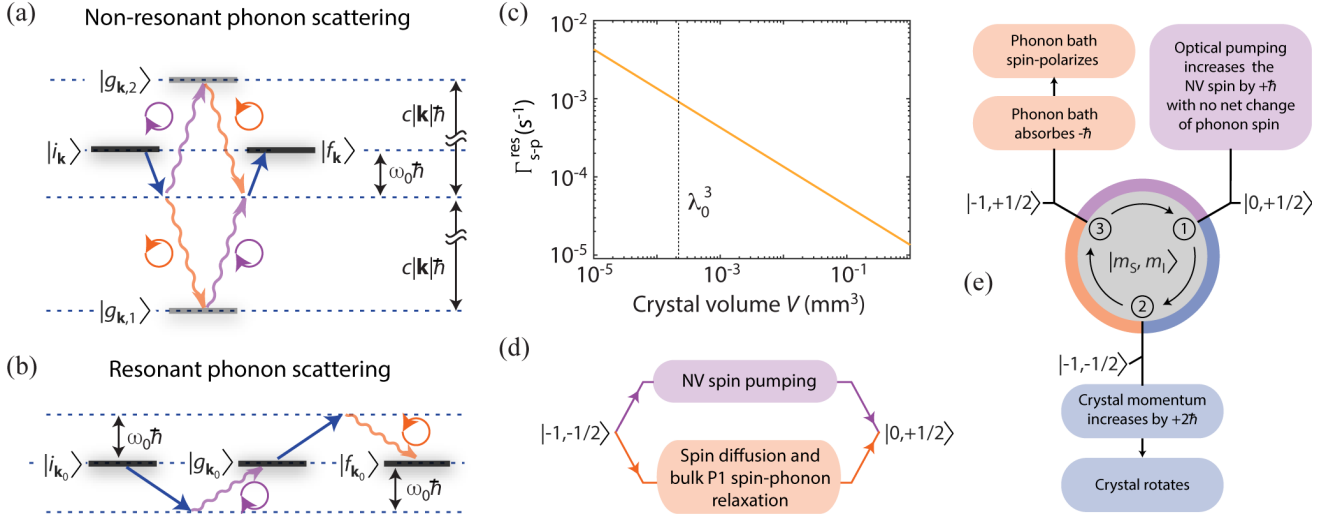

**Fig. 3: Spin-phonon angular momentum conversion. (a)** Non-resonant mechanism of spin-phonon conversion, i.e., $c|\mathbf{k}| > \omega_0$. Straight (wavy) arrows indicate spin (phonon-spin) change; clock-wise (anti-clock-wise) corresponds to phonons with negative (positive) angular momentum. The initial and final states are $|i_\mathbf{k}\rangle = |...n_{\mathbf{k}-}, n_{\mathbf{k}+},...0, +1/2\rangle$ and $|f_\mathbf{k}\rangle = |...n_{\mathbf{k}-} - 1, n_{\mathbf{k}+} + 1,...-1, -1/2\rangle$, respectively. The upper and lower virtual states are $|g_{\mathbf{k},1}\rangle = |...n_{\mathbf{k}-} - 1, n_{\mathbf{k}+},...-1, +1/2\rangle$, and $|g_{\mathbf{k},2}\rangle = |...n_{\mathbf{k}-}, n_{\mathbf{k}+} + 1,...-1, +1/2\rangle$, respectively. **(b)** Resonant mechanism, i.e., $c|\mathbf{k}_0| = \omega_0$. The notation used for all kets is the same as in (a), except that $\mathbf{k} \to \mathbf{k}_0$, and the intermediate state is $|g_{\mathbf{k}_0}\rangle = |...n_{\mathbf{k}_0-} - 1, n_{\mathbf{k}_0+},...-1, +1/2\rangle$. **(c)** Resonant spin-phonon conversion rate as a function of the crystal volume; $\lambda_0 \equiv 2\pi c/\omega_0$ denotes the wavelength of resonant phonons, not supported by crystals of smaller size. **(d)** NV–P1 spin 'reset' involving NV spin pumping and P1 spin diffusion into the bulk. Spin-phonon relaxation subsequently transfers the P1 polarization to the phonon bath. (e) Schematic representation of the NV–P1 spin cycle. Repeated sequences of NV spin pumping, P1-enabled spin cross-relaxation, and NV–P1 spin resets simultaneously produce crystal rotation and polarization of the phonon bath spin.

rate from these resonant, second-order processes is approximately given by the formula

$$\Gamma_{s-p}^{(2)} \cong \Gamma_{s-p}^{(2,res)} \approx \frac{\alpha}{10c\hbar}\left(\frac{k_B T}{\rho V}\right)^{1/2}\left(\frac{2\pi\eta}{r_{\min}^9}\right)^{1/4}, \quad (8)$$

where $k_B$ denotes Boltzmann's constant, and $T$ is the temperature. Interestingly, $\Gamma_{s-p}^{(2)}$ grows with the inverse square root of the crystal volume, implying that spin-phonon conversion is greater for diamond micro-particles. Since the wavelength of resonant phonon modes $\lambda_0 \equiv 2\pi/|\mathbf{k}_0| = 2\pi c/\omega_0$ is of order 60 μm, this mechanism is quenched in sufficiently small crystals (unable to support these phonon modes). In all cases, nonetheless, we find $\Gamma_{s-p}^{(2)} \lesssim 10^{-1}$ s$^{-1} \ll \Gamma_{s-r}$ (Fig. 3c), hence allowing us to conclude spin-phonon angular momentum conversion is not a sizable competing mechanism to spin-crystal rotation transfer.

The latter must not be interpreted, however, as implying phonons play no role in the crystal-rotation-pumping process. Phonons are key to optically repolarizing the NV center, though this process has zero net input of angular momentum into the phonon bath (see Supplementary Material, Section V, and Refs. [29,30]). On the other hand, spin-lattice relaxation of bulk P1s (which spin polarize into $|-1/2\rangle$ via spin diffusion from NV-coupled P1s) ultimately requires the transfer of (negative) angular momentum into the phonon bath. Therefore, the cycle of NV–P1 spin initialization, evolution, and reset must be viewed as one simultaneously leading to net crystal rotation and phonon-bath-spin pumping, as sketched in Figs. 3d and 3e.

## V. DISCUSSION AND OUTLOOK

While our description thus far has been limited to a single NV–P1 pair, the experimental observation of spin-to-crystal momentum conversion will likely require the use of spin ensembles. Since the dynamics of each pair is coupled to the rotation of the solid — in turn, impacting all members of the ensemble — it is natural to wonder about the conditions required to treat individual contributions to the torque on the crystal as independent from each other. To address this question, we first rewrite the Hamiltonian in Eq. (3) as (Supplementary Material, Section VI)

$$H_\phi = \frac{J_z^2}{2\mathcal{J}} + \sum_j d_{2,j}(\mathbf{r})\delta_{2-,j} + d_{2,j}^*(\mathbf{r})\delta_{2+,j}, \quad (9)$$

where $H_\phi \equiv U_\phi H U_\phi^\dagger$ with $U_\phi \equiv \exp(-i\phi(S_z + I_z)/\hbar)$, and $J_z = S_z + I_z + L_z$ is the total angular momentum (we denote $S_z \equiv \sum_j S_{z,j}$, $I_z \equiv \sum_j I_{z,j}$). The sum in Eq. (9) represents the standard (truncated) dipolar interaction of an ensemble of NV–P1 pairs in a static solid under energy matching conditions, implying that all rotation-derived effects are encapsulated in the first term. To make these effects explicit, we transform $H_\phi$ to the basis set where all terms in the sum are diagonal, i.e., where $U_\mu(d_{2,j}(\mathbf{r})\delta_{2-,j} + d_{2,j}^*(\mathbf{r})\delta_{2+,j})U_\mu^\dagger = \hbar^2 d_{2,j}'\mu_{z',j}$ with $\mu_{z',j}$ denoting a Pauli operator along a (pair-dependent) virtual axis $z'$. Limiting our description to the subspace involving states of crystal angular momentum $m_L$, the Hamiltonian takes the final form

$$H_{\phi,\mu}^{(m)} \equiv U_\mu H_\phi^{(m)} U_\mu^\dagger \approx -\hbar^2 \sum_j \left(d'_{2,j}\mu_{z',j} - (m-1)\mu_{x',j}/\mathcal{I}\right)$$
$$+(\hbar^2/\mathcal{I}) {\sum_{j\neq i}}' (\mu_{+,j}\mu_{-,i} + \mu_{-,j}\mu_{+,i}), \quad (10)$$

where we ignore constant terms, and we assume $\|d'_2\| \gg 1/\mathcal{I}$ (Supplementary Material, Section VI). In the Hamiltonian representation of Eq. (10), the first term in the upper sum can be viewed as a (local) Zeeman interaction with an effective magnetic field of amplitude proportional to the NV–P1 pair dipolar coupling, whereas the second term represents a (global) transverse field whose amplitude grows with faster crystal rotation. Finally, the primed sum (comprising only NV–P1 pairs of similar dipolar interaction strength) amounts to a rotation-induced inter-pair coupling term, independent of the inter-pair distance. Remarkably, this interaction can mediate entanglement between remote NV–P1 pairs, but because the coupling amplitude is inversely proportional to the crystal's moment of inertia, long rotational coherence lifetimes — of order $\mathcal{I}/\hbar$ — would be required to make this process observable. Under optical excitation, the system coherence time is dictated (at best) by the inverse optical pumping rate, $\Gamma_o^{-1}$, much smaller than $\mathcal{I}/\hbar$ for realistic conditions. In our present regime, therefore, we can correctly describe the impact of the ensemble on the crystal dynamics simply as a sum of independent spin-pair contributions.

Experimentally observing the interplay between spin and crystalline angular momenta can capitalize on a variety of techniques explicitly conceived to sense weak forces[31,32]. Among them, silicon-crystal double-paddle oscillators[33] — capable of detecting torques as weak as $10^{-18}$ N·m at room temperature[34] — are well-suited to the present application, because their large footprint can support mm-sized diamond crystals. For a crude comparison, we express the expected torque as

$$\tau = \frac{d\langle L_z \rangle}{dt} \sim 2\hbar \eta V \Gamma_{s-r}, \quad (11)$$

where we assume, for simplicity, near optimum NV-P1 polarization in $|0, +1/2\rangle$, a condition one can approach with reasonable illumination power densities of $\sim 1$ mW/μm$^2$ (see Supplementary Material, Section I). For crystals with moderate NV-P1 pair concentrations ($\eta \sim 5$ ppm), we find $\tau \sim 10^{-17}$ N·m for optical excitation over a $\sim 50$-μm-radius spot in a 300-μm-thick crystal. Further, the mechanically-detected spectrum that emerges — dominated by the strong P1 hyperfine interaction with its host nitrogen — serves as a signature to distinguish spin-induced torques from undesired sources (see Supplementary Material, Section VII and Ref. [35]).

In the opposite limit of diamond nano-particles[36-38], much higher detection sensitivities — from $10^{-21}$ N·m/Hz$^{1/2}$ and up to $10^{-29}$ N·m/Hz$^{1/2}$ — have been predicted[39] and demonstrated[40] using optical tweezers hence making this route also feasible; in particular, sample heating (and the ensuing NV-P1 spin energy mismatch it creates[22]) can be minimized with the use of Paul traps[41-43]. As an alternative to torque sensing, here one could capitalize on schemes adapted to detecting rotational velocities via birefringence-induced modulation of a probe laser[15]. Unlike the former, this latter strategy reveals the time integrated effect of optical excitation, and thus could help expose spin-rotation conversion in systems where the NV-P1 pair density is low.

Although our description centered on NV and P1 centers in diamond, similar derivations apply to other spin systems provided that: (*i*) one of the defects can be optically pumped (through spin-dependent optical excitation or via broadband illumination and spin-selective intersystem crossing); (*ii*) the spin numbers are different and only one has total angular momentum greater than ½ (either in the form of an orbital singlet with spin number $S \geq 1$ or an orbital doublet with $S = 1/2$ and sufficiently large spin orbit interaction); and (*iii*) both spins have suitably long lifetimes (so that they can be tuned in and out of resonance with sufficient change in flip-flop rate that the effect can be observed). Besides the NV–P1 pair discussed herein, other defect combinations in systems such as SiC or garnet materials appear plausible.

Extending the ideas introduced herein promises intriguing opportunities in various uncharted fronts. For example, unlike present schemes to inducing rotation, the ability to initialize and manipulate paramagnetic centers provides a versatile handle to control the rotational dynamics of the host crystal, which could be exploited to investigate the limits of quantum superposition in mesoscopic systems. Provided the rotational coherence of the host crystal is sufficiently long, it will also be interesting to investigate the impact of rotation on the collective dynamics of the spin ensemble, which, perhaps, could lead to forms of 'coherence protection' akin to that observed in heterogeneous ensembles of oscillators confined to an optical cavity[44,45].

Along the same lines, the interplay between spin-lattice relaxation and chiral phonons — here found comparatively inefficient at the NV–P1 pair level — could nonetheless be exploited at the single defect level. One possibility could be to mechanically pump the NV (and/or P1) spin, for instance, by stimulating spin-polarized acoustic phonons matched to the spin resonance frequency. To this end, one could resort to existing photo-acoustic methods based on timed femtosecond laser pulses[46,47], in this case tailored so as to coherently inject chiral phonons into the diamond lattice.

**ACKNOWLEDGEMENTS**

C.A.M. thanks Erwin Hahn, Dimitris Sakellariou, Andreas Trabesinger, and Alexander Pines for stimulating his interest on the topic through early discussions, and, more recently, Eugene Chudnovsky for fruitful conversations on the ideas herein. P.R.Z. and C.A.M. acknowledge support from the National Science Foundation through grants NSF-1619896, NSF-1903839, and NSF-1914945, and from Research Corporation for Science Advancement through a FRED Award; they also acknowledge access to the facilities and research infrastructure of the NSF CREST IDEALS, grant number NSF-HRD-1547830. A.A.W. acknowledges support from the University of Melbourne ECR mobility fellowship scheme. M.W.D. acknowledges support from the Australian Research Council (DE170100169).

**APPENDIX A: METHODS**

***Dynamics using the Trotter-Suzuki decomposition.*** The tight-binding representation in Fig. 2(a) corresponds to the unitary dynamics given by the Hamiltonian in Eq. (3) and the non-unitary processes $\Gamma_o$ and $\Gamma_L$ (optical pumping and rotational dephasing, respectively). The standard Trotterization allows for a stepwise evolution, where the system is evolved in small time steps $\Delta t$. Here, $\Delta t$ is much smaller than the shortest time-scale in the problem, including $\Gamma_d$, $\Gamma_o$ and $\Gamma_L$. The projection due to optical pumping (and the reset of the P1 spin state) follows the standard Quantum Jump recipe[48]. In practice, this implies a stochastic projection of population from state $|-1,-1/2\rangle$ to the state $|0,+1/2\rangle$. The dephasing $\Gamma_L$ corresponds to the analog Quantum Drift procedure[49], and consists in a stochastic randomization of the phase of each state $|m_L\rangle$. The time dependence of the observable (in this case, the probability density associated to the wave-function) is obtained after averaging trajectories.

# Supplemental material

# Mechanical rotation via optical pumping of paramagnetic impurities

Pablo R. Zangara[1], Alexander Wood[3], Marcus W. Doherty[4], Carlos. A. Meriles[1,2,†]

[1]Dept. of Physics, CUNY-City College of New York, New York, NY 10031, USA. [2]CUNY-Graduate Center, New York, NY 10016, USA. [3]Centre for Quantum Computation and Communication Technology, School of Physics, The University of Melbourne, Melbourne, Victoria 3010, Australia. [4]Laser Physics Centre, Research School of Physics and Engineering, Australian National University, Canberra, Australian Capital Territory 0200, Australia. [†]E-mail: cmeriles@ccny.cuny.edu.

## I. Estimates for angular momentum pumping

The torque generated in the crystal is a function of the optical pumping rate $\Gamma_o$, the spin-spin interaction rate $\Gamma_d$, the dephasing rate $\Gamma_L$, the NV relaxation rate $\Gamma_{NV}$, and the P1 relaxation rate $\Gamma_{P1}$. In the special case where the optical pumping is synchronous with the dipolar (double-) spin-flip, i.e. when $\Gamma_d \sim \Gamma_o$, we can compute the spin-rotation conversion rate $\Gamma_{s-r}$, using the Fermi golden rule (FGR):

$$\Gamma_{s-r} = \frac{2\pi}{\hbar^2} \overline{|\langle m_L + 2, -1, -1/2|H_d|m_L, 0, +1/2\rangle|^2} \rho_{ss}(\omega_{NV} - \omega_{P1}), \quad (S1)$$

where $\rho_{ss}(\omega_{NV} - \omega_{P1})$ is the total density of spin states, and the overbar denotes an ensemble average over the spatial distribution of NV-P1 pairs. The matrix element in the previous formula is non-zero due to the term $d_2(\mathbf{r})\lambda_+^2 \delta_{2-} + d_2^*(\mathbf{r})\lambda_-^2 \delta_{2+}$ in $H_d$. Thus, we need to compute $\overline{|d_2(\mathbf{r})|^2}$,

$$\overline{|d_2(\mathbf{r})|^2} = \eta \int_{r_{min}}^{r_{max}} r^2 dr \int_0^\pi \sin(\theta)\, d\theta \int_0^{2\pi} d\varphi \frac{9\alpha^2}{16 r^6} \sin^4 \theta$$

$$= \frac{2\pi}{5} \eta \alpha^2 \left( \frac{1}{r_{min}^3} - \frac{1}{r_{max}^3} \right). \quad (S2)$$

Here, $\eta$ is the number density of P1 centers such that there is only one center per volume $2\pi r_{max}$ on average, and $r_{min} \sim 1$ nm is the smallest allowed distance between the NV and P1 centers (below 1 nm, the electronic wavefunctions of these defects begin to overlap and so they behave as a collective system rather than individual spin systems). Then,

$$\Gamma_{s-r} = \frac{4\pi^2}{5\hbar^2} \eta \alpha^2 \left( \frac{1}{r_{min}^3} - \frac{1}{r_{max}^3} \right) \rho_{ss}(\omega_{NV} - \omega_{P1}). \quad (S3)$$

It is convenient to interpret $\rho_{ss}(\omega_{NV} - \omega_{P1})$ as a lineshape factor accounting for the broadening of the NV-P1 spin resonance. This broadening is naturally given by the spin-spin coupling (dipolar P1-P1 interaction), which, in turn, is quantified by $\Gamma_d = \hbar \overline{|d_2(\mathbf{r})|^2}/(2\pi)$. Dephasing of the rotational states $|m_L\rangle$ at a rate given by $\Gamma_L$ produces an additional contribution to the lineshape factor. If much greater than $\Gamma_d$, it ultimately quenches the spin-rotation mechanism as shown in Fig. 2(c) in the main text. Note that a similar argument can be used for the optical illumination, which tends to pin the NV in the $|m_S = 0\rangle$ state if too intense. Then,

$$\Gamma_{s-r} = \frac{4\pi^2}{5\hbar^2} \eta \alpha^2 \left( \frac{1}{r_{min}^3} - \frac{1}{r_{max}^3} \right) \rho_{ss}(\Gamma_d, \Gamma_o, \Gamma_L) \approx \frac{\Gamma_d^3}{\left(\Gamma_d^2 + \Gamma_L^2\right)^{1/2} \left(\Gamma_d^2 + \Gamma_o^2\right)^{1/2}}, \quad (S4)$$

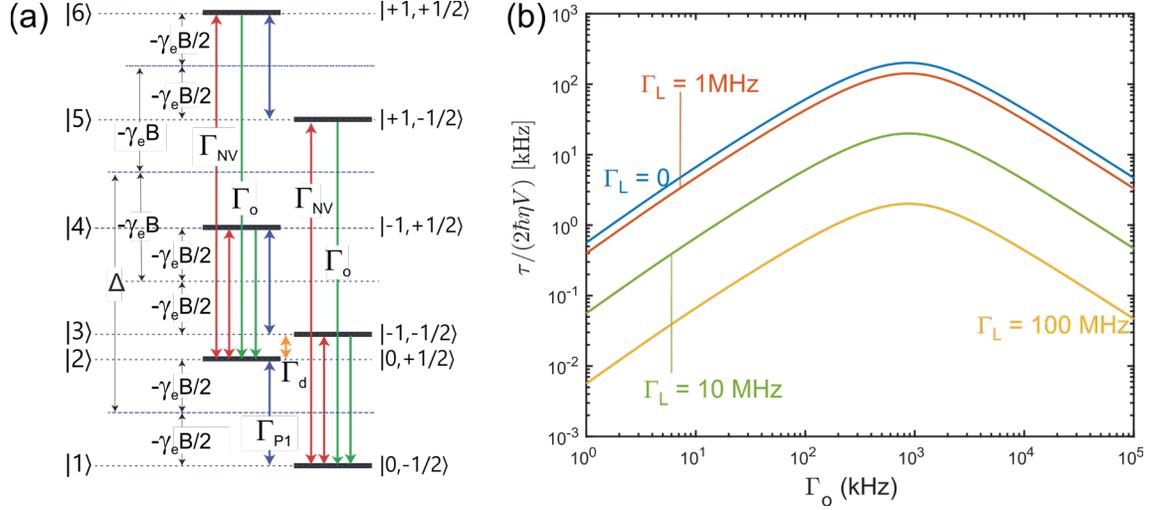

**Figure S1:** (a) NV-P1 energy diagram; close to the matching condition, where states $|2\rangle$ and $|3\rangle$ are nearly degenerate. Green arrows indicate optical pumping at a rate $\Gamma_o$. Red arrows denote NV spin-lattice relaxation at rates $\Gamma_{NV}$; blue arrows denote the P1 spin diffusion rate $\Gamma_{P1}$; NV–P1 spin cross-relaxation takes place at a rate $\Gamma_d$ (orange arrow). (b) Normalized torque as a function of the optical pumping rate for various rotational decoherence rates $\Gamma_L$.

where the last approximate formula reproduces the dependence of the slopes in Fig. 2(c) (main text) as a function of $\Gamma_L$ for fixed $\Gamma_d$ and $\Gamma_o$.

Equation (S4) indicates excessive optical excitation is detrimental to the generation of crystal rotation because it pins the NV in $|0\rangle$, thus precluding momentum transfer to the crystal. To calculate the average torque as a function of the NV spin optical pumping rate, we first determine the steady-state spin populations for the NV–P1 set. To this end, we write

$$\frac{d\mathbf{n}}{dt} = \mathbf{\Gamma} \cdot \mathbf{n} = 0, \qquad (S5)$$

where $\mathbf{n} = (n_1, \ldots n_6)$, with the components $n_i$, $i = 1 \ldots 6$ denoting the probabilities of the NV–P1 spin system occupying the $i$-th level so that $\sum_{i=1}^{6} n_i = 1$. To solve Eq. (S5), we express the relaxation matrix $\mathbf{\Gamma}$ in terms of the NV spin optical pumping rate $\Gamma_o$, the NV–P1 dipolar coupling $\Gamma_d$, and the NV and P1 spin lattice relaxation rates $\Gamma_{NV}$ and $\Gamma_{P1}$, respectively (see Fig. S1(**a**)). A more accurate assessment of $\Gamma_{P1}$ should be given not by the raw $T_1$ time, but by the $P1 - P1$ spin diffusion time, which in turn implies $\Gamma_{P1} \approx \Gamma_d$. A numerical solution of Eq. (S5) allows us to determine the population difference $\delta n_{23} = n_2 - n_3$ as a function of the NV spin optical pumping rate $\Gamma_o$, and thus calculate the torque induced in the crystal as (see Fig. S1(**b**))

$$\tau = \frac{d\langle L_z \rangle}{dt} \sim 2\hbar\eta V \Gamma_{s-r}(n_2 - n_3). \qquad (S6)$$

When writing Eq. (S6), we first note that P1 spin-lattice relaxation can only affect the net spin of the phonon bath but leaves the crystal angular momentum unchanged [1]; further, we assume optical pumping of the NV spin has no impact on the net angular momentum of the phonon bath or crystal (Section V), and that all NV-P1 pairs act independently (Section VI).

## II. Interaction with spin-polarized phonons

Phonons are introduced in our model by allowing the point defects to undergo small displacements $\boldsymbol{\delta R}$ from their equilibrium coordinates $\mathbf{r}$:

$$\mathbf{r} \rightarrow \mathbf{r} + \boldsymbol{\delta R}, \tag{S7}$$

where $\mathbf{r} = \mathbf{r}_{\mathrm{NV}} - \mathbf{r}_{\mathrm{P1}}$ is the inter-spin vector. A first order expansion of the dipolar Hamiltonian on the displacements yields

$$H_{\mathrm{d}}(\mathbf{r} + \boldsymbol{\delta R}) \approx H_{\mathrm{d}}(\mathbf{r}) + \sum_{\xi} \frac{\partial}{\partial \delta R_{\xi}} [H_{\mathrm{d}}(\mathbf{r} + \boldsymbol{\delta R})]_{\delta R=0} \delta R_{\xi} \tag{S8}$$

where $\xi = x, y, z$. The zero-order term is the spin-rotation interaction $H_{\mathrm{d}}$ introduced in Eq. (2) in the main text and corresponds to the rigid-rotator. The first-order term — incorporating into the model the lattice elasticity — can be cast as the sum of spin-phonon and spin-rotation-phonon interactions, here denoted as $H_{\mathrm{s-p}}$ and $H_{\mathrm{s-r-p}}$, respectively. In what follows (Sections II, III, IV), we consider only the spin-phonon term $H_{\mathrm{s-p}}$, since this is the leading first order contribution. Note that in this limit, the azimuthal coordinate collapses to $\varphi$, as crystal rotations (described via $\phi$) can be ignored.

An explicit evaluation of the spatial derivatives yields

$$\begin{aligned}
H_{\mathrm{s-p}} = \frac{1}{2} e^{i\varphi} &\big[ \big( g_{1-r}\sin(\theta) + ig_{1-\varphi} + g_{1-\theta}\cos(\theta) \big) \boldsymbol{\delta R}_{\mathrm{x}} \\
&+ \big( -ig_{1-r}\sin(\theta) + g_{1-\varphi} - ig_{1-\theta}\cos(\theta) \big) \boldsymbol{\delta R}_{\mathrm{y}} \big] \\
+ \frac{1}{2} e^{-i\varphi} &\big[ \big( g_{1+r}\sin(\theta) - ig_{1+\varphi} + g_{1+\theta}\cos(\theta) \big) \boldsymbol{\delta R}_{\mathrm{x}} \\
&+ \big( ig_{1+r}\sin(\theta) + g_{1+\varphi} + ig_{1+\theta}\cos(\theta) \big) \boldsymbol{\delta R}_{\mathrm{y}} \big] \\
+ &\big( g_{0,r}\cos(\theta) - g_{0,\theta}\sin(\theta) \big) \boldsymbol{\delta R}_{\mathrm{z}}
\end{aligned} \tag{S9}$$

with the definitions

$$\vec{\nabla}[d_0(\mathbf{r})]\delta_0 = g_{0,r}\hat{\mathbf{r}} + g_{0,\varphi}\hat{\boldsymbol{\varphi}} + g_{0,\theta}\hat{\boldsymbol{\theta}}$$

$$\vec{\nabla}[d_1(\mathbf{r})]\delta_{1-} = g_{1+r}\hat{\mathbf{r}} + g_{1+\varphi}\hat{\boldsymbol{\varphi}} + g_{1+\theta}\hat{\boldsymbol{\theta}}$$

$$\vec{\nabla}[d_1^*(\mathbf{r})]\delta_{1+} = g_{1-r}\hat{\mathbf{r}} + g_{1-\varphi}\hat{\boldsymbol{\varphi}} + g_{1-\theta}\hat{\boldsymbol{\theta}}$$

$$\vec{\nabla}[d_2(\mathbf{r})]\delta_{2-} = g_{2+r}\hat{\mathbf{r}} + g_{2+\varphi}\hat{\boldsymbol{\varphi}} + g_{2+\theta}\hat{\boldsymbol{\theta}}$$

$$\vec{\nabla}[d_2^*(\mathbf{r})]\delta_{2+} = g_{2-r}\hat{\mathbf{r}} + g_{2-\varphi}\hat{\boldsymbol{\varphi}} + g_{2-\theta}\hat{\boldsymbol{\theta}}$$

We now perform a transformation from the spherical coordinate vectors $(\hat{\mathbf{r}}, \hat{\boldsymbol{\varphi}}, \hat{\boldsymbol{\theta}})$ to the Cartesian set $(\hat{\mathbf{e}}_x, \hat{\mathbf{e}}_y, \hat{\mathbf{e}}_z)$. Additionally, by introducing $\boldsymbol{\delta R}_\pm = \boldsymbol{\delta R}_x \pm i\boldsymbol{\delta R}_y$ we rewrite $H_{\mathrm{s-p}}$ as

$$H_{\mathrm{s-p}} = b_0 \boldsymbol{\delta R}_z \delta_0 + b_1 [\boldsymbol{\delta R}_- \delta_{1+} + \boldsymbol{\delta R}_+ \delta_{1-}], \tag{S10}$$

where

$$b_0 = -\frac{3\alpha}{2r^4}\cos(\theta)(1 - 5\cos 2\theta), \tag{S11}$$

$$b_1 = -\frac{3\alpha}{16r^4}(3\cos\theta + 4\cos 3\theta). \tag{S12}$$

Since $\boldsymbol{\delta R}_\xi = \boldsymbol{\delta R}_{NV,\xi} - \boldsymbol{\delta R}_{P1,\xi}$ with $\xi = x, y, z$, we rewrite the displacements of each individual spin $j = $ NV, P1 as

$$\boldsymbol{\delta R}_{j,\xi} = \sum_{\mathbf{k},p} \left[\frac{\hbar}{2\rho V \omega_{\mathbf{k}p}}\right]^{1/2} \hat{u}_{\mathbf{k},p,\xi} \left(e^{i\mathbf{k}\cdot\mathbf{r}_j} a_{\mathbf{k},p} + e^{-i\mathbf{k}\cdot\mathbf{r}_j} a_{\mathbf{k},p}^\dagger\right), \quad (S13)$$

where $V$ and $\rho$ respectively denote the crystal volume and density, $\mathbf{k}$ and p are respectively the phonon wavevector and (Cartesian-) polarization vector, $\hat{u}_{\mathbf{k},p}$ is the unit phonon displacement vector, $a_{\mathbf{k},p}^\dagger$ and $a_{\mathbf{k},p}$ are the phonon creation and annihilation operators, and $\omega_{\mathbf{k}p}$ is the phonon frequency.

Now we write $\boldsymbol{\delta R}_\pm$ as

$$\boldsymbol{\delta R}_\pm = \sum_{\mathbf{k},p} \left[\frac{\hbar}{2\rho V \omega_{\mathbf{k}p}}\right]^{1/2} \left(\hat{u}_{\mathbf{k},p,x} \pm i\hat{u}_{\mathbf{k},p,y}\right) \left[\left(e^{i\mathbf{k}\cdot\mathbf{r}_{P1}} - e^{i\mathbf{k}\cdot\mathbf{r}_{NV}}\right) a_{\mathbf{k},p} \right. \\ \left. + \left(e^{-i\mathbf{k}\cdot\mathbf{r}_{P1}} - e^{-i\mathbf{k}\cdot\mathbf{r}_{NV}}\right) a_{\mathbf{k},p}^\dagger\right] \quad (S14)$$

and a linear expansion assuming $\mathbf{k}\cdot\mathbf{r}_{NV}, \mathbf{k}\cdot\mathbf{r}_{P1} \ll 1$ yields

$$\boldsymbol{\delta R}_\pm \approx \sum_{\mathbf{k},p} i\mathbf{k}\cdot\mathbf{r} \left[\frac{\hbar}{2\rho V \omega_{\mathbf{k}p}}\right]^{\frac{1}{2}} \left(\hat{u}_{\mathbf{k},p,x} \pm i\hat{u}_{\mathbf{k},p,y}\right) \left[a_{\mathbf{k},p} - a_{\mathbf{k},p}^\dagger\right] \quad (S15)$$

Circularly polarized phonons are introduced according to the standard definition for photons [2,3],

$$a_{\mathbf{k},\pm}^\dagger = \mp \frac{\left(a_{\mathbf{k},x}^\dagger \pm i a_{\mathbf{k},y}^\dagger\right)}{\sqrt{2}} \quad (S16)$$

$$a_{\mathbf{k},\pm} = \mp \frac{\left(a_{\mathbf{k},x} \mp i a_{\mathbf{k},y}\right)}{\sqrt{2}} \quad (S17)$$

This definition is consistent with the accepted convention of positive helicity ($+\hbar$ angular momentum) corresponding to left-circular polarization (LCP), and negative helicity ($-\hbar$ angular momentum) corresponding to right-circular polarization (RCP).

Then,

$$H_{s-p} \approx \sum_{\mathbf{k}} i\mathbf{k}\cdot\mathbf{r} \{b_0 \pi_{\mathbf{k},0} \delta_0 + b_1 \pi_{\mathbf{k},1-} \delta_{1+} + b_1 \pi_{\mathbf{k},1+} \delta_{1-}\} \quad (S18)$$

where

$$\pi_{\mathbf{k},1\pm} = \pm \left[\frac{\hbar}{2\rho V \omega_\mathbf{k}}\right]^{\frac{1}{2}} \left(a_{\mathbf{k},\mp} + a_{\mathbf{k},\pm}^\dagger\right) \quad (S19)$$

$$\pi_{\mathbf{k},0} = \left[\frac{\hbar}{2\rho V \omega_\mathbf{k}}\right]^{\frac{1}{2}} \left(a_{\mathbf{k},z} - a_{\mathbf{k},z}^\dagger\right) \quad (S20)$$

The operator $\pi_{\mathbf{k},1+}$ either *creates* a phonon with linear momentum $\mathbf{k}$ and $+\hbar$ angular momentum, or *destroys* a phonon with linear momentum $\mathbf{k}$ and $-\hbar$ angular momentum. Reciprocally, the operator $\pi_{\mathbf{k},1-}$ *creates* a phonon with linear momentum $\mathbf{k}$ and $-\hbar$ angular momentum, or *destroys* a phonon with linear momentum $\mathbf{k}$ and $+\hbar$ angular momentum. Global conservation of angular momentum is then clear in Eq. (S10) since every spin flip is accompanied by a corresponding gain or loss of spin angular momentum in the phonon bath.

### III. Conversion to spin-polarized phonons: Fermi golden rule

There is no one-phonon transition between $|0, +1/2\rangle$ and $|-1, -1/2\rangle$ since there are no spin operators in $H_{s-p}$ that raise or lower both spins at once. So, the lowest order rate driven by $H_{s-p}$ is a two phonon Raman-type elastic scattering that couples an initial state $|i_\mathbf{k}\rangle = |\ldots n_{\mathbf{k},-}, n_{\mathbf{k},+}, \ldots, 0, +1/2\rangle$ to a final state $|f_\mathbf{k}\rangle = |\ldots n_{\mathbf{k},-} - 1, n_{\mathbf{k},+} + 1, \ldots, -1, -1/2\rangle$. Here, $n_{\mathbf{k},-}$ and $n_{\mathbf{k},+}$ represent the phonon-spin-polarized populations with a wave vector $\mathbf{k}$. The decrease in angular momentum from the NV–P1 spin cross-relaxation is compensated by the annihilation of one RCP phonon and the creation of one LCP phonon. The intermediate state connecting $|i_\mathbf{k}\rangle$ and $|f_\mathbf{k}\rangle$ can be either $|g_{\mathbf{k},1}\rangle = |\ldots n_{\mathbf{k},-} - 1, n_{\mathbf{k},+}, \ldots, -1, +1/2\rangle$ or $|g_{\mathbf{k},2}\rangle = |\ldots n_{\mathbf{k},-}, n_{\mathbf{k},+} + 1, \ldots, -1, +1/2\rangle$ (intermediate states of the form $|g_{\mathbf{k},3}\rangle = |\ldots n_{\mathbf{k},-} - 1, n_{\mathbf{k},+}, \ldots, 0, -1/2\rangle$ and $|g_{\mathbf{k},4}\rangle = |\ldots n_{\mathbf{k},-}, n_{\mathbf{k},+} + 1, \ldots, 0, -1/2\rangle$ are forbidden since the operator $\pi_{\mathbf{k},1+}\delta_{1-}$ — required to act twice — cannot drive the transition $|0, +1/2\rangle \leftrightarrow |0, -1/2\rangle$). Also note that conservation of the linear momentum ($\mathbf{k}$) is implicit in the choice of the initial, intermediate, and final states.

The FGR formula for second-order processes is given by

$$\Gamma_{s-p}^{(2)} = \frac{2\pi}{\hbar^2} \overline{\left|\sum_{g(\mathbf{k})} \frac{\langle f_\mathbf{k}|H_{s-p}|g\rangle\langle g|H_{s-p}|i_\mathbf{k}\rangle}{E_i - E_g}\right|^2} \rho_{ss}(\omega_{NV} - \omega_{P1}), \qquad (S21)$$

where $E_i$ and $E_g$ represent the corresponding energies of the initial and intermediate states, respectively. The overbar denotes an ensemble average over the NV-P1 pairs (spatial distribution) and momentum $\mathbf{k}$.

We first evaluate the second order matrix elements,

$$\sum_{g(\mathbf{k})} \frac{\langle f_\mathbf{k}|H_{s-p}|g\rangle\langle g|H_{s-p}|i_\mathbf{k}\rangle}{E_i - E_g} = (\mathbf{k} \cdot \mathbf{r})^2 b_1^2 \frac{\omega_0 \sqrt{n_{\mathbf{k},-}(n_{\mathbf{k},+} + 1)}}{2\rho V \omega_\mathbf{k}(\omega_0^2 - \omega_\mathbf{k}^2)}, \qquad (S22)$$

where $\omega_0 \equiv |\gamma_e|B$. In what follows, we assume $n_{\mathbf{k},+}$ and $n_{\mathbf{k},-}$ are independent of the phonon polarization; in addition, we use the low-frequency approximation $\omega_\mathbf{k} = c|\mathbf{k}|$, with $c$ denoting the speed of sound in diamond (assumed isotropic). Since,

$$n_\mathbf{k} = \left[\exp\left(\frac{\hbar c k}{k_B T}\right) - 1\right]^{-1} \qquad (S23)$$

the square modulus of the second order matrix element is

$$\left|\sum_{g(\mathbf{k})} \frac{\langle f_\mathbf{k}|H_{s-p}|g\rangle\langle g|H_{s-p}|i_\mathbf{k}\rangle}{E_i - E_g}\right|^2 = (\hat{\mathbf{k}} \cdot \mathbf{r})^4 b_1^4 \frac{\omega_0^2 \omega_\mathbf{k}^2 n_\mathbf{k}(n_\mathbf{k} + 1)}{4\rho^2 V^2 c^4 (\omega_0^2 - \omega_\mathbf{k}^2)^2}. \qquad (S24)$$

Computing now the average over the spatial distribution and momentum $\mathbf{k}$, we obtain.

$$\left|\sum_{g(\mathbf{k})} \frac{\langle f_{\mathbf{k}}|H_{\text{s-p}}|g\rangle\langle g|H_{\text{s-p}}|\iota_{\mathbf{k}}\rangle}{E_\iota - E_g}\right|^2$$

$$= \left(\eta \int_{r_{min}}^{r_{max}} r^2 dr \int_0^\pi \sin(\theta_r)\, d\theta_r \int_0^{2\pi} d\varphi_r\right) \times$$

$$\times \left(\frac{V}{(2\pi)^3}\int_0^\infty k^2 dk \int_0^\pi \sin(\theta_k)\, d\theta_k \int_0^{2\pi} d\varphi_k\right)(\hat{\mathbf{k}}\cdot\mathbf{r})^4 b_1^4 \frac{\omega_0^2 \omega_{\mathbf{k}}^2 n_{\mathbf{k}}(n_{\mathbf{k}}+1)}{4\rho^2 V^2 c^4 (\omega_0^2 - \omega_{\mathbf{k}}^2)^2}$$

$$= \left(\eta \int_{r_{min}}^{r_{max}} r^2 dr \int_0^\pi \sin(\theta_r)\, d\theta_r \int_0^{2\pi} d\varphi_r\right) \times$$

$$\times \left(\int_0^\pi \sin(\theta_k)\, d\theta_k \int_0^{2\pi} d\varphi_k\right)\frac{(\hat{\mathbf{k}}\cdot\mathbf{r})^4 b_1^4 \omega_0^2}{2^5 \pi^3 \rho^2 V c^7} \int_0^\infty d\omega\, \frac{\omega^4 n_\omega (n_\omega+1)}{(\omega_0^2 - \omega^2)^2}.$$

In particular,

$$\int_0^\pi \sin(\theta_k)\, d\theta_k \int_0^{2\pi} d\varphi_k (\hat{\mathbf{k}}\cdot\mathbf{r})^4 = \pi r^4 \left(\frac{24}{15}(\cos\theta_r)^2(\sin\theta_r)^2 + \frac{4}{5}(\cos\theta_r)^4 + \frac{4}{5}(\sin\theta_r)^4\right),$$

and

$$\int_0^\pi \sin(\theta_r)\, d\theta_r \int_0^{2\pi} d\varphi_r\, (\cos\theta_r)^2(\sin\theta_r)^2(3\cos\theta_r + 4\cos 3\theta_r)^4 \approx 30\pi,$$

$$\int_0^\pi \sin(\theta_r)\, d\theta_r \int_0^{2\pi} d\varphi_r\, [(\cos\theta_r)^4 + (\sin\theta_r)^4](3\cos\theta_r + 4\cos 3\theta_r)^4 \approx 376\pi.$$

Then,

$$\left|\sum_{g(\mathbf{k})} \frac{\langle f_{\mathbf{k}}|H_{\text{s-p}}|g\rangle\langle g|H_{\text{s-p}}|\iota_{\mathbf{k}}\rangle}{E_\iota - E_g}\right|^2$$

$$\approx \eta\left(\frac{1}{r_{min}^9} - \frac{1}{r_{max}^9}\right)\frac{\alpha^4 \omega_0^2}{2^5 \pi \rho^2 V c^7}\frac{1}{21}\int_0^\infty d\omega\, \frac{\omega^4 n_\omega(n_\omega+1)}{(\omega_0^2 - \omega^2)^2}. \quad (S25)$$

We can now rewrite Eq. (S21) as

$$\Gamma_{\text{s-p}}^{(2)} = \frac{2\pi}{\hbar^2}\frac{\Lambda \omega_0^2}{\rho^2 V c^7}\int_0^\infty d\omega\, \frac{\omega^4 n_\omega(n_\omega+1)}{(\omega_0^2 - \omega^2)^2}\rho_{ss}(\omega_{NV} - \omega_{P1}), \quad (S26)$$

where we defined the factor

$$\Lambda \approx \eta\left(\frac{1}{r_{min}^9} - \frac{1}{r_{max}^9}\right)\frac{\alpha^4}{2^5 \pi}\frac{1}{21} \approx \frac{\eta}{r_{min}^9}\left(\frac{\alpha^4}{5}\right)\frac{1}{2^7 \pi}. \quad (S27)$$

The integral over $\omega$ can be written in terms of unit-less variables by introducing $\nu = \hbar\omega/k_B T$,

$$\int_0^\infty d\omega\, \frac{\omega^4 n_\omega(n_\omega+1)}{(\omega_0^2 - \omega^2)^2} = \frac{k_B T}{\hbar}\int_0^\infty d\nu\, \frac{\nu^4 n_\nu(n_\nu+1)}{(\nu_0^2 - \nu^2)^2}, \quad (S28)$$

where $\nu_0 = \hbar\omega_0/k_B T \approx 2.36\times 10^{-4}$ (and we assume $T = 293$ K). Since the integrand in the previous expression diverges when $\nu \to \nu_0$, a crude way to overcome the singularity relies on adding the natural lineshape $\Gamma_d$ of spin-resonance. In such a case,

$$\int_0^\infty d\omega \frac{\omega^4 n_\omega(n_\omega+1)}{(\omega_0^2-\omega^2)^2} \approx \frac{k_B T}{\hbar} \int_0^\infty d\nu \frac{\nu^4 n_\nu(n_\nu+1)}{(\nu_0^2-\nu^2)^2+\nu_1^4}, \quad (S29)$$

where $\nu_1 = \hbar\Gamma_d/k_B T = 2.6\times 10^{-8}$. In calculating Eq. (S29) we use a conservative estimate $\Gamma_d \sim 1$ MHz. This regularization yields

$$\int_0^\infty d\omega \frac{\omega^4 n_\omega(n_\omega+1)}{(\omega_0^2-\omega^2)^2+\Gamma_d^4} \approx \frac{k_B T}{\hbar} 1.4\times 10^{19}, \quad (S30)$$

which we can now use to compute the rate $\Gamma_{s-p}^{(2)}$ as a function of the crystal volume $V$. In Fig. S2(**a**) we explicitly show such a relation, assuming $\eta = 1$ ppm, $r_{min} = 1$ nm, $T = 293$ K, and $\rho_{ss}(\omega_{NV} - \omega_{P1}) \sim 1/\Gamma_d$. We find $\Gamma_{s-p}^{(2)} \ll \Gamma_{s-r}$ over a broad range of rotor decoherence rates $\Gamma_L$, hence suggesting that conversion of spin polarization from the NV–P1 pair into phonon spin is comparatively inefficient.

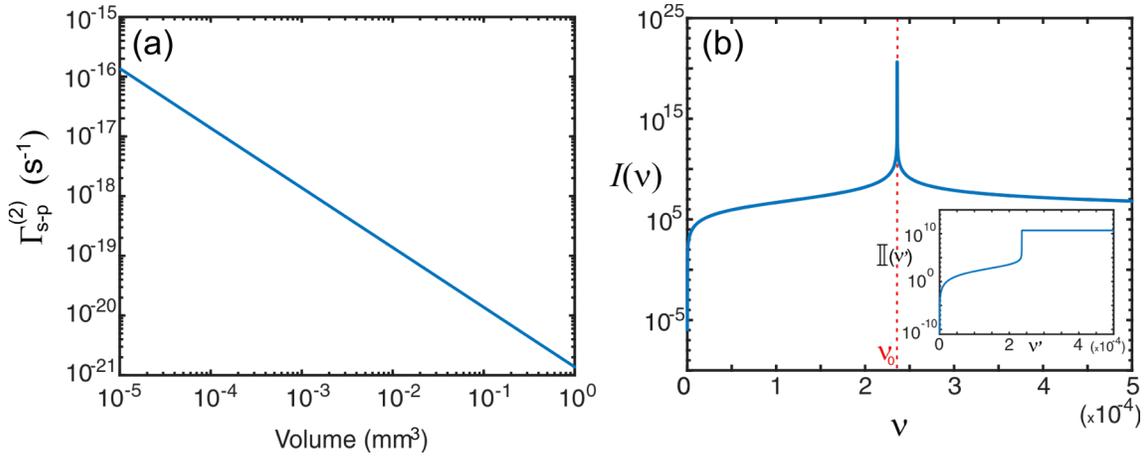

**Figure S2**. (**a**) Non-resonant spin/phonon-spin conversion rate $\Gamma_{s-p}^{(2)}$ as a function of the crystal volume $V$ as derived from Eqs. (S26) and (S30). (**b**) The integrand in Eq. (S29) as a function of the unit-less variable $\nu$. Inset: partial integral $\mathbb{I}(\nu') = \int_0^{\nu'} I(\nu)d\nu$; saturation above $\nu_0$ indicates the largest contribution to $\Gamma_{s-p}^{(2)}$ is due to near-resonance phonons.

## IV. Spin-polarized phonons: resonant case

Before disregarding transfer to the phonon spin bath as a mechanism for NV–P1 relaxation, it is worth re-examining Eq. (S26) with focus on the resonant case, where the phonon frequency matches the Zeeman splitting, i.e., where $\omega \approx \omega_0$. We can better assess the relative weight of these contributions in Fig. S2(**b**), where we plot the renormalized integrand in Eq. (S29)

$$I(\nu) = \frac{\nu^4 n_\nu(n_\nu+1)}{(\nu_0^2-\nu^2)^2+\nu_1^4}, \quad (S31)$$

revealing a large peak near resonance. Upon computing the partial integral $\mathbb{I}(\nu') = \int_0^{\nu'} I(\nu)\,d\nu$ (insert to Fig. S2(**b**)), we conclude that resonant or near-resonant contributions — not properly taken into account by Eq. (S24) — are dominant, implying this type of phonon scattering process must be analyzed carefully.

We start by defining the resonant condition for phonons,

$$\omega_{\mathbf{k}_0} = c|\mathbf{k}_0| = \omega_0, \qquad (S32)$$

where, for simplicity, we limit our discussion to the case where the magnetic field is ~51 mT; for future reference, we note that resonant phonons share a wavelength $\lambda_0 = 2\pi/|\mathbf{k}_0| \approx 60$ μm.

The resonant condition implies a triple degeneracy involving three states:

$$|i_{\mathbf{k}_0}\rangle = |\ldots n_{\mathbf{k}_0,-}, n_{\mathbf{k}_0,+}, \ldots, 0, +1/2\rangle$$

$$|f_{\mathbf{k}_0}\rangle = |\ldots n_{\mathbf{k}_0,-} - 1, n_{\mathbf{k}_0,+} + 1, \ldots, -1, -1/2\rangle$$

$$|g_{\mathbf{k}_0}\rangle = |\ldots n_{\mathbf{k}_0,-} - 1, n_{\mathbf{k}_0,+}, \ldots, -1, +1/2\rangle$$

Unlike the case in the prior section, such three-level system does not require a virtual process to induce transitions from $|i_{\mathbf{k}_0}\rangle$ to $|f_{\mathbf{k}_0}\rangle$, since there is no energy difference to be paid in the intermediate hopping. First, we compute the coupling matrix elements:

$$\langle g_{\mathbf{k}_0}|H_{\text{s-p}}|i_{\mathbf{k}_0}\rangle = \sum_{\mathbf{k}} i\mathbf{k}\cdot\mathbf{r}\, b_1 \sqrt{\frac{\hbar}{2\rho V \omega_k}} \langle g_{\mathbf{k}_0}|a_{\mathbf{k},-}S_- I_z|i_{\mathbf{k}_0}\rangle$$

$$= \sum_{\mathbf{k}} i\mathbf{k}\cdot\mathbf{r}\, b_1 \sqrt{\frac{\hbar}{2\rho V \omega_k}} \frac{\sqrt{n_{\mathbf{k}_0,-}}}{2}\delta(\mathbf{k}, \mathbf{k}_0)$$

$$= i\mathbf{k}_0\cdot\mathbf{r}\,\frac{b_1}{2}\sqrt{\frac{\hbar}{2\rho V \omega_0 \left(\exp\left(\frac{\hbar\omega_0}{k_B T}\right) - 1\right)}}.$$

We can further simplify the previous expression by assuming $\exp(\hbar\omega_0/k_B T) \approx 1 + \hbar\omega_0/k_B T$,

$$\langle g_{\mathbf{k}_0}|H_{\text{s-p}}|i_{\mathbf{k}_0}\rangle \approx i\hat{\mathbf{k}}_0\cdot\mathbf{r}\,\frac{b_1}{2c}\sqrt{\frac{k_B T}{2\rho V}}. \qquad (S33)$$

Analogously,

$$\langle f_{\mathbf{k}_0}|H_{\text{s-p}}|g_{\mathbf{k}_0}\rangle = \sum_{\mathbf{k}} i\mathbf{k}\cdot\mathbf{r}\, b_1 \sqrt{\frac{\hbar}{2\rho V \omega_k}} \langle f_{\mathbf{k}_0}|a^\dagger_{\mathbf{k},+}S_z I_-|g_{\mathbf{k}_0}\rangle$$

$$= -\sum_{\mathbf{k}} i\mathbf{k}\cdot\mathbf{r}\, b_1 \sqrt{\frac{\hbar}{2\rho V \omega_k}} \sqrt{n_{\mathbf{k}_0,+} + 1}\,\delta(\mathbf{k}, \mathbf{k}_0)$$

$$\approx -i\hat{\mathbf{k}}_0\cdot\mathbf{r}\,\frac{b_1}{c}\sqrt{\frac{k_B T}{2\rho V}}. \qquad (S34)$$

The Hamiltonian in this subspace is then given by

$$H_{\text{s-p}}^{[0]} = \begin{array}{c|ccc} & |i_{\mathbf{k}_0}\rangle & |g_{\mathbf{k}_0}\rangle & |f_{\mathbf{k}_0}\rangle \\ \hline \langle i_{\mathbf{k}_0}| & E & G & 0 \\ \langle g_{\mathbf{k}_0}| & G^\dagger & E & 2G \\ \langle f_{\mathbf{k}_0}| & 0 & 2G^\dagger & E \end{array}$$

where $E$ is the energy of the subspace (which can be set to zero) and

$$G = -i\hat{\mathbf{k}}_0 \cdot \mathbf{r} \frac{b_1}{2c}\sqrt{\frac{k_B T}{2\rho V}}. \tag{S35}$$

Ultimately, we seek to attain a characteristic time-scale for the correlation function

$$P_{f,i}(t) = \left|\langle f_{\mathbf{k}_0}|exp\left\{-\frac{iH_{s-p}^{[0]}t}{\hbar}\right\}|i_{\mathbf{k}_0}\rangle\right|^2. \tag{S36}$$

From the exact solution of the Schrodinger equation, we have

$$P_{f,i}(t) = \frac{2^2}{5^2}\left[1 - \cos\left(\frac{\sqrt{5}|G|t}{\hbar}\right)\right]^2, \tag{S37}$$

and a short-time expansion yields

$$P_{f,i}(t) \approx \frac{2^2}{5^2}\left[1 - \left(1 - \frac{\frac{5}{2}|G|^2 t^2}{\hbar^2}\right)\right]^2 \approx (\hat{\mathbf{k}}_0 \cdot \mathbf{r})^4 \frac{b_1^4 (k_B T)^2}{2^4 c^4 \rho^2 V^2 \hbar^4} t^4. \tag{S38}$$

Averaging over the ensemble of NV–P1 pairs and over all directions of $\hat{\mathbf{k}}_0$, we find

$$\overline{P_{f,i}(t)} \approx \left(\eta \int_{r_{min}}^{r_{max}} r^2 dr \int_0^\pi \sin(\theta_r)\, d\theta_r \int_0^{2\pi} d\varphi_r\right) \times \left(\frac{1}{4\pi}\int_0^\pi \sin(\theta_k)\, d\theta_k \int_0^{2\pi} d\varphi_k\right)$$
$$\times (\hat{\mathbf{k}}_0 \cdot \mathbf{r})^4 \frac{b_1^4 (k_B T)^2}{2^4 c^4 \rho^2 V^2 \hbar^4} t^4, \tag{S39}$$

and after some algebra,

$$\overline{P_{f,i}(t)} \approx \frac{172\alpha^4 (k_B T)^2}{2^6 \pi c^4 \rho^2 V^2 \hbar^4} t^4 \eta \int_{r_{min}}^{r_{max}} r^2 dr \left(\frac{3}{16 r^4}\right)^4 2\pi^2 r^4$$
$$\approx \frac{\alpha^4 (k_B T)^2}{2^6 \pi c^4 \rho^2 V^2 \hbar^4} t^4 \eta \left(\frac{1}{r_{min}^9} - \frac{1}{r_{max}^9}\right)\pi^2 \frac{1}{21}$$
$$\approx \frac{\pi^2 (k_B T)^2 \Lambda}{2 c^4 \rho^2 V^2 \hbar^4} t^4. \tag{S40}$$

Rewriting Eq. (S40) as $\overline{P_{f,i}(t)} \approx (t/\tau_{s-p}^{(res)})^4$, we find the characteristic phonon-spin interaction time $\tau_{s-p}^{(res)}$ and corresponding relaxation rate

$$\Gamma_{s-p}^{(2,res)} \equiv \left(\tau_{s-p}^{(res)}\right)^{-1} = \left(\frac{4.8}{13120.3}\right)^{\frac{1}{4}} \frac{\alpha}{c\hbar}\left(\frac{k_B T}{\rho V}\right)^{\frac{1}{2}}\left(\frac{2\pi\eta}{r_{min}^9}\right)^{\frac{1}{4}} \approx \frac{\alpha}{10 c\hbar}\left(\frac{k_B T}{\rho V}\right)^{\frac{1}{2}}\left(\frac{2\pi\eta}{r_{min}^9}\right)^{\frac{1}{4}}. \tag{S41}$$

This estimate is explicitly shown as a function of $V$ in Fig. 3(c) of the main text.

## V. Angular momentum conservation in the optical cycle of the NV center

A key assumption in our analysis is that the optical pumping of the NV spin has no impact on the net angular momentum of the phonon bath or crystal. In order to address this issue, we first identify the three systems involved in the optical cycle of the NV: (1) the light field, (2) the NV center's electrons, and (3) the lattice phonon field. Each system has orbital and spin angular momenta. However, in the following we will restrict the light field to only possessing spin angular momentum. There are two specific questions that we need to consider: (1) how is angular momentum exchanged between the systems during the optical cycle? and (2) what orbital phonon angular momentum is generated? The latter question is important to assessing whether optical cycling can induce a rotational angular momentum of the diamond that either reinforces or opposes the rotational angular momentum we are seeking to generate by spin-spin cross-relaxations.

The NV center's optical cycle has many alternate radiative and non-radiative decay pathways. We will restrict the discussion to just the four dominant pathways: (1) direct radiative decay from $^3E$ to $^3A_2$, (2) a non-radiative transition with $^3E$ followed by radiative decay to $^3A_2$, (3) non-radiative decay from the $A_1$ state of the $^3E$ level to the $m_S = 0$ state of the $^3A_2$, and (4) non-radiative decay from the $E_\pm$ state of the $^3E$ level to the $m_s = 0$ state of the $^3A_2$. The first two pathways are spin-conserving. The second two pathways flip the spin from $m_S = \pm 1$ to $m_S = 0$ and are responsible for the NV center's optical spin polarization and readout. Note that the non-radiative transition in (2) is the well-established dynamic Jahn-Teller effect in the $^3E$ [4], whilst the pathways (3) and (4) are the upper intersystem crossings (ISCs) identified in [5].

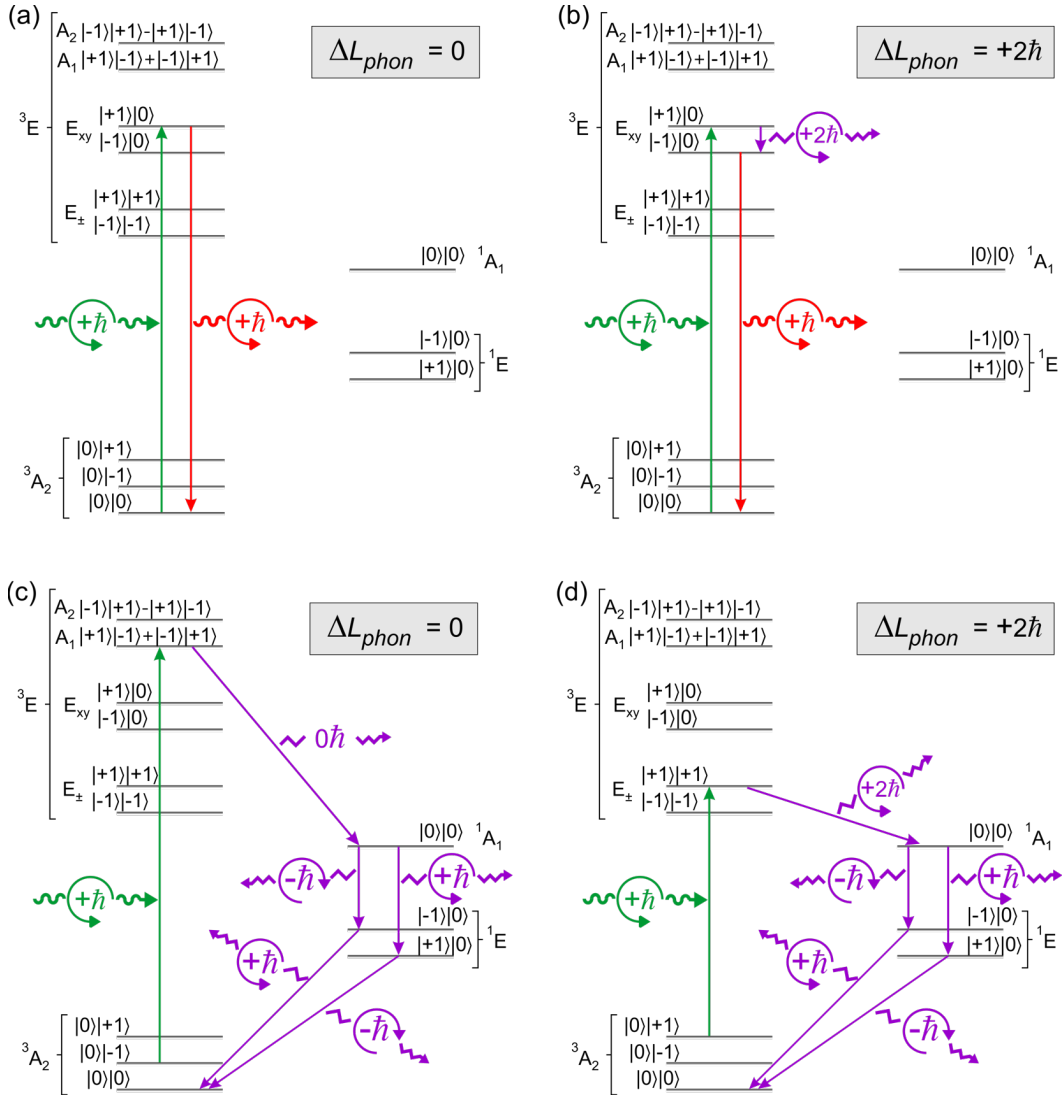

**Figure S3.** Angular momentum conservation in the four dominant optical cycle pathways of the NV center under LCP excitation, respectively involving: (a) direct radiative decay from the $^3$E to $^3$A$_2$, (b) a non-radiative transition with the $^3$E followed by radiative decay to the $^3$A$_2$, (c) non-radiative decay from the A$_1$ state of the $^3$E level to the $m_s = 0$ state of the $^3$A$_2$, and (d) non-radiative decay from the E$_\pm$ state of the $^3$E level to the $m_s = 0$ state of the $^3$A$_2$. The radiative decay pathways of (a) and (b) are independent of spin and are thus the same for the $m_s = \pm 1$ and $m_s = 0$ spin states. Accordingly, (a) and (b) only depict them for the $m_s = 0$ states. Transitions are denoted by arrows. Photons and phonons are denoted by wavy and zig-zag arrows, respectively. Each electronic state is labelled by their total orbital and spin state projections $|m_L\rangle|m_s\rangle$. $\Delta L_{\text{phon}}$ is the net change in phonon field angular momentum.

In Fig. S3, we sketch the optical cycles involving each of the above pathways for the particular case of left circularly polarized (LCP) excitation, and evaluate the net change in phonon angular momentum. Inverse results are obtained for right circularly polarized (RCP) excitation. We see that pathways (1) and (3) do not induce net change in phonon angular momentum, whilst the other two pathways generate $+2\hbar$ ($-2\hbar$) of phonon angular momentum for LCP (RCP) excitation. The origin of the phonon angular momentum generation in (2) and (4) is the phonon-induced transitions between the $|+\rangle$ and $|-\rangle$ orbital angular momentum states of the $^3E$ level. As these transitions involve the creation/annihilation of a single phonon, the phonon mode must have $2\hbar$ of angular momentum. This could be purely orbital angular momentum or orbital and spin angular momentum. Further exploration of this phenomenon is required to determine what forms the angular momentum takes.

The key conclusions from the figures are that: (A) the excitation photon spin angular momentum is conserved by conversion into a combination of emitted photon, electron spin and phonon angular momentum; and (B) if off-resonance linearly polarized light, containing equal LCP and RCP components, is used to excite the NV center, then on average there will be no net change in the phonon angular momentum. Hence, under these excitation conditions, optical cycling will not influence the rotational angular momentum generated by spin-spin cross-relaxation.

## VI. Rotator-mediated pair-to-pair interactions

To properly gauge rotor-driven correlations between remote NV–P1 pairs within the same crystal, we rewrite the Hamiltonian in Eq. (3) as,

$$H = \frac{L_z^2}{2\mathcal{J}} + d_{2(a)}(\mathbf{r})\lambda_+^2 \delta_{2-}^{(a)} + d_{2(a)}^*(\mathbf{r})\lambda_-^2 \delta_{2+}^{(a)} + d_{2(b)}(\mathbf{r})\lambda_+^2 \delta_{2-}^{(b)} + d_{2(b)}^*(\mathbf{r})\lambda_-^2 \delta_{2+}^{(b)} \quad (S42)$$

where the indexes $a$, $b$ label each of the pairs. It is important to stress that the generalization to an arbitrary number of interacting pairs is quite straightforward.

We introduce now the transformation operator,

$$U_\phi = exp\left\{-i\phi \left(S_z^{(a)} + I_z^{(a)} + S_z^{(b)} + I_z^{(b)}\right)/\hbar\right\}, \quad (S43)$$

where $S_z^{(a)}$ denotes the z-component of the spin operator for the NV in pair $(a)$, $I_z^{(a)}$ denotes the z-component of the spin operator for the P1 in pair $(a)$, and we use equivalent definitions for pair $(b)$.

In order to compute $H_\phi = U_\phi H U_\phi^\dagger$ we start with a few auxiliary identities. First, we have

$$U_\phi \delta_{2-}^{(a)} U_\phi^\dagger = exp\left\{-i\phi S_z^{(a)}/\hbar\right\} S_-^{(a)} exp\left\{+i\phi S_z^{(a)}/\hbar\right\} exp\left\{-i\phi I_z^{(a)}/\hbar\right\} I_-^{(a)} exp\left\{+i\phi I_z^{(a)}/\hbar\right\}$$

$$= exp\{+i\phi\} S_-^{(a)} exp\{+i\phi\} I_-^{(a)}$$

$$= exp\{+2i\phi\}\delta_{2-}^{(a)}, \tag{S44}$$

and

$$U_\phi \delta_{2+}^{(a)} U_\phi^\dagger = exp\left\{-i\phi S_z^{(a)}/\hbar\right\} S_+^{(a)} exp\left\{+i\phi S_z^{(a)}/\hbar\right\} exp\left\{-i\phi I_z^{(a)}/\hbar\right\} I_+^{(a)} exp\left\{+i\phi I_z^{(a)}/\hbar\right\}$$

$$= exp\{-i\phi\}S_+^{(a)} exp\{-i\phi\}I_+^{(a)}$$

$$= exp\{-2i\phi\}\delta_{2+}^{(a)}. \tag{S45}$$

Additionally,

$$U_\phi L_z^2 U_\phi^\dagger |\psi\rangle = -\hbar^2 U_\phi \frac{\partial^2}{\partial \phi^2} U_\phi^\dagger |\psi\rangle$$

$$= -\hbar^2 U_\phi \left[\left(\frac{\partial^2}{\partial \phi^2} U_\phi^\dagger\right)|\psi\rangle + 2\left(\frac{\partial}{\partial \phi} U_\phi^\dagger\right)\left(\frac{\partial}{\partial \phi}|\psi\rangle\right) + U_\phi^\dagger \left(\frac{\partial^2}{\partial \phi^2}|\psi\rangle\right)\right]$$

$$= \left[\left(S_z^{(a)} + I_z^{(a)} + S_z^{(b)} + I_z^{(b)}\right)^2 + 2\left(S_z^{(a)} + I_z^{(a)} + S_z^{(b)} + I_z^{(b)}\right)L_z + L_z^2\right]|\psi\rangle$$

$$= \left[\left(S_z^{(a)} + I_z^{(a)} + S_z^{(b)} + I_z^{(b)}\right) + L_z\right]^2 |\psi\rangle. \tag{S46}$$

Then,

$$H_\phi = U_\phi H U_\phi^\dagger$$

$$= \frac{1}{2\mathcal{J}}\left[\left(S_z^{(a)} + I_z^{(a)} + S_z^{(b)} + I_z^{(b)}\right) + L_z\right]^2 + d_{2(a)}\delta_{2-}^{(a)} + d_{2(a)}^*\delta_{2+}^{(a)} + d_{2(b)}\delta_{2-}^{(b)}$$
$$+ d_{2(b)}^*\delta_{2+}^{(b)}. \tag{S47}$$

Since $\hbar^2|d_{2(a,b)}| \gg \hbar^2/\mathcal{J}$, the leading energy scale in $H_\phi$ is given by the dipolar spin flip terms, hence suggesting a natural basis where these terms are diagonal. We therefore define the pair of transformations $U_\mu^{(a)}$ and $U_\mu^{(b)}$ such that

$$U_\mu^{(a)}\left(d_{2(a)}\delta_{2-}^{(a)} + d_{2(a)}^*\delta_{2+}^{(a)}\right)U_\mu^{(a)\dagger} = \hbar^2 d_{2(a)}' \mu_{z',(a)} \tag{S48}$$

$$U_\mu^{(b)}\left(d_{2(b)}\delta_{2-}^{(b)} + d_{2(b)}^*\delta_{2+}^{(b)}\right)U_\mu^{(b)\dagger} = \hbar^2 d_{2(b)}' \mu_{z',(b)} \tag{S49}$$

Here, $\mu_{z',(a,b)}$ is a Pauli matrix relative to a virtual axis $z'$, which in turn depends on the specific pair ($a$ or $b$). The global transformation is defined as $U_\mu = U_\mu^{(a)} \otimes U_\mu^{(b)}$, with an obvious generalization to an arbitrary number of pairs.

Using the auxiliary identities

$$U_\mu \left(S_z^{(a)} + I_z^{(a)}\right) U_\mu^\dagger = -\frac{\hbar}{2}\mathbf{I} - \hbar\left(\mu_{+',(a)} + \mu_{-',(a)}\right), \tag{S50}$$

$$U_\mu \left(S_z^{(a)} + I_z^{(a)}\right)^2 U_\mu^\dagger = \frac{5\hbar^2}{4}\mathbf{I} + \hbar^2\left(\mu_{+',(a)} + \mu_{-',(a)}\right), \tag{S51}$$

with equivalent versions for pair $b$, and after some algebra,

$$H_{\phi,\mu}^{(m)} = U_\mu H_\phi U_\mu^\dagger = \hbar^2\left(d_{2(a)}'\mu_{z',(a)} + d_{2(b)}'\mu_{z',(b)}\right) + \frac{\hbar^2}{2\mathcal{J}}(m^2 - 2m + 3)\mathbf{I}$$

$$+\frac{\hbar^2}{\mathcal{J}}(1-m)\big(\mu_{+',(a)} + \mu_{-',(a)} + \mu_{+',(b)} + \mu_{-',(b)}\big)$$
$$+\frac{\hbar^2}{\mathcal{J}}\big(\mu_{+',(a)}\mu_{+',(b)} + \mu_{+',(a)}\mu_{-',(b)} + \mu_{-',(a)}\mu_{+',(b)} + \mu_{-',(a)}\mu_{-',(b)}\big) \quad (S52)$$

From the above expression, we conclude the impact of crystal rotation on the dynamics of NV–P1 pairs can be seen as that of an effective 'transverse' field acting separately on each spin pair (third term in Eq. (S52)) and a coupling term between separate, otherwise non-interacting pairs (fourth term in Eq. (S52)). Equation (10) in the main text corresponds to a secularized version of Eq. (S52), where terms of the form $\mu_{+',(a)}\mu_{+',(b)}$ and $\mu_{-',(a)}\mu_{-',(b)}$ have been neglected.

## VII. Towards an experimental realization

Experimentally observing a torque from dipolar cross relaxation as discussed herein can leverage existing techniques based on high-$Q$ oscillators explicitly conceived to sense weak forces [6,7,8]. An important concern is the spatial scale of the proposed experiment, with smaller configurations suited to intrinsically more sensitive devices, but with concomitantly lower sample volumes (and torques) and higher resonant frequencies. For concreteness, we analyze here one possible route using a macroscopic silicon-crystal double-paddle oscillator (DPO) capable of detecting torques as weak as $10^{-18}$ N·m at room temperature [9].

Silicon DPOs are well-suited to the present application because their large footprint — exceeding 1 cm$^2$ — can support a commercially available diamond crystal and the ~kHz mechanical resonance frequencies typical in this class of oscillators are slow compared with the cross-relaxation transfer rates. We consider a diamond containing nitrogen concentrations of ~1 ppm mounted on the DPO as depicted in Fig. S4(**a**). Optical pumping with 532 nm light polarizes the NV centers, and for an applied magnetic field parallel to the NV axes and satisfying the NV-P1 energy-matching criteria, dipolar cross-relaxation excites the torsional mode depicted in Fig. S4(**a**). An optical lever composed of an additional laser and position sensitive detector measures the deflection of the oscillator paddle and thus the induced torque.

Assuming an NV–P1 pair concentration of $\eta \sim 10^{18}$ cm$^{-3}$ (the equivalent of ~5 ppm), an optical polarization rate $\Gamma_o = 100$ kHz, and a small dephasing $\Gamma_L \lesssim 1$ MHz, Eq. (S6) predicts a unit volume torque of $\tau/V \sim 2\times 10^{-5}$ N·m$^{-2}$. Therefore, for a focal spot of radius $\varsigma = 50$ μm traversing through a ~300-μm-thick diamond crystal, the net torque amounts to $\tau \sim 10^{-17}$ N·m. To estimate the required laser power, we write as a crude approximation $\Gamma_o \approx \Gamma_o^{(\text{sat})}\big(1 - \exp(-\Upsilon/\Upsilon^{(\text{sat})})\big)$, where $\Gamma_o^{(\text{sat})} = 1$ MHz, $\Upsilon$ denotes the illumination intensity, and $\Upsilon^{(\text{sat})} = 1$ mW/μm$^2$. For the conditions assumed herein, $\Gamma_o \sim \Gamma_o^{(\text{sat})} \Upsilon/\Upsilon^{(\text{sat})}$ and thus the laser power can be estimated as $P = \pi\varsigma^2 \Upsilon^{(\text{sat})} \Gamma_o/\Gamma_o^{(\text{sat})} \sim 780$ mW. This level of optical intensity is not dissimilar from what is used in previous demonstrations of ultrasensitive ensemble NV magnetometry [10].

It would be thus possible to implement detection protocols where the laser is turned on and off synchronously with the resonator oscillation to excite a torque. Further, given the large hyperfine couplings of the P1 with its nitrogen host (of order ~100 MHz), a magnetic field sweep should yield a discernible, characteristic collection of NV-P1 resonances [11-13]; thus, the mechanically-detected spectrum that emerges (Fig. S4(**c**)) serves as a signature to distinguish spin-induced torques from other sources, including the laser beam itself. While our proposed scheme is considerably simplified, it motivates further technical study into the prospects for macroscopic

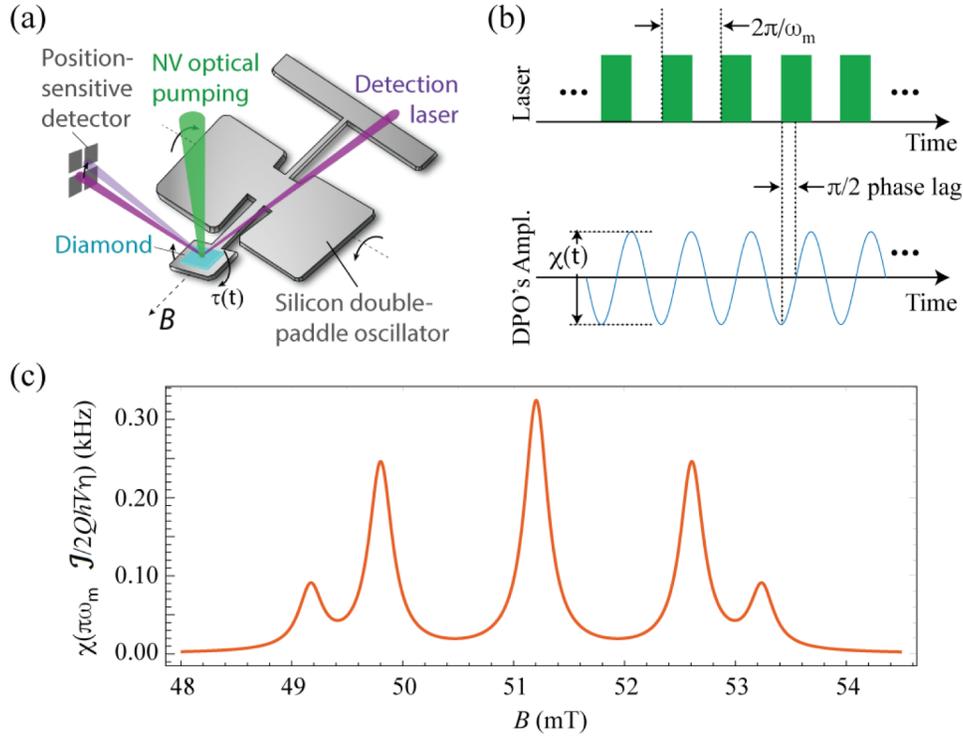

**Figure S4:** (a) Conversion of spin polarization into mechanical rotation of the diamond crystal could be attained with the help of a high-$Q$ double-paddle torsional oscillator. (b) Detection protocol. Optical excitation takes the form of a train of light pulses at the oscillator resonance frequency. (c) Calculated amplitude of the oscillator motion as a function of the externally applied magnetic field.

sensing schemes to be applied.

In the opposite limit, small diamond particles and nanodiamonds potentially offer platforms with much higher detection sensitivities of up to $10^{-21}$ N·m/Hz$^{1/2}$ as recently reported [14]; a sensitivity of up to $10^{-29}$ N·m/Hz$^{1/2}$ has been predicted [15] using optically trapped nanodiamonds in ultra-high vacuum. Sample heating limits the vacuum pressure attainable and thus the resultant $Q$-factor of the trapped-nanodiamond torsional resonator; further, laser-induced sample heating in any proposed scheme would also induce an NV-P1 energy mismatch. Recent experimental work using electrically trapped nanodiamonds in ion traps [16] may offer an alternative torque sensing platform, in addition to other proposed means of investigating and harnessing spin-rotation coupling at the nanoscale [17].